\documentclass[12pt]{article}
\usepackage{appendix}
\usepackage{mathtools}
\usepackage{amsmath}
\usepackage{amsthm}
\usepackage{amsfonts}
\usepackage{amssymb}
\usepackage{commath}
\usepackage{biblatex}
\usepackage{tikz}
\usepackage{algorithm2e}
\usepackage{caption}
\usepackage{subcaption}
\usepackage{listings}
\usepackage[margin=1in]{geometry}

\renewcommand{\P}{\mathbb{P}}

\newcommand{\erf}{\text{erf}}
\newcommand{\Ei}{\text{Ei}}

\author{Gavin Ridley, Benoit Forget, Timothy Burke}
\title{Resonance Scattering Treatment with the Windowed Multipole Formalism}
\date{\today}

\begin{document}
\maketitle

\section{Abstract}

A new method for directly sampling the resonance upscattering effect is presented.  Alternatives have relied on inefficient rejection sampling techniques or large tabular storage of relative velocities. None of these approaches, which require pointwise energy data, are particularly well suited to the windowed multipole cross section representation. The new method called multipole analytic resonance scattering (MARS) overcomes these limitations by inverse transform sampling from the target relative velocity distribution where the cross section is expressed in the multipole formalism. The closed form relative speed distribution contains a novel special
function we deem the incomplete Faddeeva function, and we present the first results on
its efficient numerical evaluation.

\section{Introduction}
Early continuous energy Monte Carlo neutron transport programs sampled scattering
from nuclei in thermal motion assuming that the scattering cross section is effectively
constant within the scattering kernel \cite{gelbardEpithermalScatteringVIM1979}.
However, as \cite{rothensteinTwobodyKineticsTreatment1995, ouisloumenModelNeutronScattering1991}
detail , the resulting scattering kernel implied by the constant cross section
approximation may be far from the actual double-differential cross section near a scattering resonance. As shown in \cite{leeImpact238UResonance2009,moriComparisonResonanceElastic2009},
the resulting error tends to cause a worst-case 11\% underestimation of the Doppler feedback coefficient in a PWR,
with even larger discrepancies in HTGR problems.

As exhibited by the PRAGMA project \cite{choiRelativeSpeedTabulation2021,choiOptimizationNeutronTracking2021}, the conventional
methods for treating this effect leave something to be desired on graphics processing unit
(GPU) architectures, which constitute the majority of computational power on today's
leading supercomputers. Due to the unique architecture of the GPU, algorithmic modifications
to standard Monte Carlo algorithms for neutron tracking can tangibly accelerate computation
\cite{choiOptimizationNeutronTracking2021}. In the same direction, we herein present a GPU-friendly method
for handling resonance upscatter when the windowed multipole (WMP) \cite{joseyWindowedMultipoleCross2016} formalism is employed to represent
cross sections. In particular, the heuristic for fast GPU code is to avoid rejection
sampling and accesses to distantly spaced places in memory, which the new method achieves.
To give context to the new method, we first recall some of the conventional methods
for modeling resonance upscattering.

The Doppler Broadening Rejection Correction (DBRC) \cite{beckerProofImplementationStochastic2009} was one of the early proposed techniques to treat the effect of
strong variations of the interaction cross section within the energetic vicinity of a scattering neutron, whereas $S(\alpha, \beta)$ tables had been used prior \cite{daganUseTablesNuclides2005}.
The method has since been implemented in numerous continuous energy Monte Carlo neutron transport programs
\cite{beckerProofImplementationStochastic2009,zoiaDopplerBroadeningNeutron2013,trumbullEffectsApplyingDoppler2013,hartImplementationDopplerBroadening2013}, and has shown to successfully model the resonance
upscatter effect. However, DBRC suffers from rejection probabilities as high as 99.995\% \cite{romanoImprovedTargetVelocity2018} for neutron energies
in the vicinity of a resonance.

The weight correction method (WCM) \cite{leeImpact238UResonance2009} also successfully models
the effect of resonances on the double-differential cross section of nuclei in thermal motion.
This method adjusts the weight of particles to provide numerically correct results even when
the constant cross section double-differential free gas distribution is employed.
WCM carries the same benefit of our newly proposed method of not requiring an additional rejection loop or tables; however, adjustments to the particle weights introduce substantial variance 
to the overall Monte Carlo simulation, thus degrading estimates on quantities of interest
\cite{choiRelativeSpeedTabulation2021}.

Another technique known as target motion sampling (TMS) \cite{viitanenExplicitTreatmentThermal2012} can be used to model the
resonance upscatter effect. However, this method is only applicable to Monte Carlo neutron transport programs employing the delta
tracking technique. Unfortunately, performance of delta tracking appears to be lackluster on GPUs \cite{rowlandDeltatrackingGPUacceleratedWARP2017}.

The relative velocity sampling (RVS) method \cite{walshAcceleratedSamplingFree2014,romanoImprovedTargetVelocity2018} was created to ameliorate the high rejection
rates characteristic to the rejection algorithms used to model resonance upscatter. These schemes, in essence, sample probability
distributions proportional to $f(x) g(x)$, where $f(x)$ is a distribution and $g(x) \in [0, 1]$. One then samples from $f(x)$ and
accepts the sample with probability $g(x)$. The RVS method moves the direct sampling from the thermal motion term to the cross section
term, thus worsening the average case rejection rate but massively improving the worst-case rejection rate.


The relative speed tabulation (RST) method \cite{choiRelativeSpeedTabulation2021} was developed to address the rejection sampling
performance impact on GPUs.
RST is the first resonance upscatter modeling technique
not requiring a rejection loop or bivariate scattering distribution tables. The key observation underlying RST is that the target
velocity distribution can be factorized into a marginal relative speed distribution encapsulating the information about the resonances
and a simple distribution of the target polar angle conditioned on the target relative speed. Consequently, the univariate relative speed
cumulative distribution can be tabulated in select areas of the pointwise cross section, and the conditional polar angle distribution
is then directly sampled without requiring any additional data or rejection step.

Despite its simplicity and efficacy on GPUs, the RST method comes with some clear disadvantages. Gigabytes of additional memory are used
in storing the relative speed cumulative distributions (CDFs) at each energy point and temperature on a pointwise cross section representation, if all nuclides have the resonance upscatter effect treated. The resonance influence
on the double differential cross section is thus restricted for practical reasons to a select few nuclides in the problem to avoid extreme
memory usage. Moreover, the method introduces some discretization error in temperature, although this was shown to be a reasonable approximation.

If one could avoid pre-tabulated relative speed distributions, this could substantially reduce the memory needs and avoid costly divergent memory accesses. Reducing serialized memory accesses on GPUs typically leads to significant speedups.

We propose a new method which does just that, and achieves this using
the windowed multipole (WMP) cross section representation \cite{joseyWindowedMultipoleCross2016}. Our new method introduces
a novel special function we have deemed the incomplete Faddeeva function which encodes the behavior of the temperature-dependent
influence of resonances on the double differential scattering distribution. It relies on the numerical inversion of the analytic
representation of the relative speed distribution under the WMP formalism, and polar angle sampling in the same manner as the RST method, but without any precomputed tables.

This contrasts the WMP-based target velocity sampling technique presented in \cite{liangTargetVelocitySampling2018},
which is similar in nature to the newly presented method in this work. However, \cite{liangTargetVelocitySampling2018}'s method relies on the separation of the target velocity
distribution into a zero kelvin cross section component and a Maxwell-Boltzmann component. This method thus
requires a numerical inversion step of the integrated scattering cross section function wrapped in a rejection loop,
similar to \cite{romanoImprovedTargetVelocity2018} but instead using a functional representation of the integrated cross section rather than tabular.

The algorithm presented in \cite{biondoAlgorithmFreeGas2021} shows how the relative speed can be sampled
in the windowed multipole framework similarly to our work. However, \cite{biondoAlgorithmFreeGas2021}
relies on fitting a sum of Gaussians to replace the poles in Eq. \ref{eq:xs_form_full}, thus representing
the zero kelvin scattering cross section in the form:
\begin{equation}
  \sigma(E) = \frac{1}{E} \sum_k \sum_j \left[ h_{s,k,j} e^{\left(\sqrt{E}-u_k\right)^2/w^2_{s,k,j}} +
    h_{a,k,j} e^{\left(\sqrt{E}-u_k\right)^2/w^2_{a,k,j}}
    \right]
\end{equation}
While it remains to be seen that Gaussians can be used to approximate all
poles appearing in a windowed multipole library, this proposed approximation introduces a
considerable number of degrees of freedom to the problem, with eighteen unknowns for each pole. The optimization
problem thus encountered is highly nonlinear and nonconvex leading to fitting difficulties. Additionally, scattering kernels in \cite{biondoAlgorithmFreeGas2021}'s formalism cannot be straightforwardly differentiated with
respect to windowed multipole parameters which would be needed to perform sensitivity analysis. 

Our new method, multipole analytic resonance scattering (MARS), only requires the same windowed multipole data as would be used in a calculation without any treatment
of the resonance upscatter effect, avoiding the need for additional tables or fitting steps. 
We demonstrate the new method's negligible performance overhead compared to other methods
for sampling the resonance upscatter effect on CPU architectures, with future work exploring
its optimized implementation on GPUs.



\section{Theory}

Rothenstein et. al. \cite{rothensteinTwobodyKineticsTreatment1995} expounds the rigorous Doppler broadened double-differential cross section. Since the exact expression for the lab frame scattering
distribution is quite complicated, authors presenting algorithms to model it typically choose to forgo the lab frame expression and instead
reason in terms of joint distributions of the target velocity and direction cosine relative to the direction
of projectile motion. After sampling the target speed and direction cosine, standard two-body collision kinematics for elastic scattering are employed, where the center of mass angular
distribution comes from the nuclear data file. The resulting physics matches the complicated
expressions of \cite{rothensteinTwobodyKineticsTreatment1995}.

Similarly considering the distribution of collision target velocities, this joint distribution is:
\begin{equation}
  f(V, \mu) = C v_r \sigma(v_r) M(T, V) f'(\mu)
  \label{eq:vmu_distr}
\end{equation}
where $M$ is the Maxwell-Boltzmann distribution of target speeds:
\begin{equation}
  M(T, V) = \frac{4}{\sqrt{\pi}} \beta^3 V^2 e^{-\beta^2 V^3}
\end{equation}
and the variable with units of inverse velocity $\beta = \sqrt{\frac{A}{2kT}}$ parameterizes the target velocities,
and the distribution $f'(\mu)$ is a uniform distribution between -1 and 1. The other variables
are $C$, the distribution's normalizing constant; $v_r$, the relative speed of the neutron with respect to the target; $\sigma$, the zero kelvin scattering cross section; $A$, the target mass; $k$, the Boltzmann constant; $T$, the absolute temperature;
and $V$, the target speed.

Classically, the approximation that
$\sigma(v_r)$ is constant has been employed. However, it was shown in \cite{ouisloumenModelNeutronScattering1991, leeImpact238UResonance2009} that this approximation
is incorrect in the vicinity of resonances, where $\sigma(v_r)$ varies over a few orders
of magnitude, preferentially causing scattering with targets of relative velocity more closely matching the scattering resonance peaks.
The various aforementioned techniques are all just methods for sampling from the distribution of Eq. \ref{eq:vmu_distr} with arbitrary forms
of the function $\sigma(v_r)$.

More specific knowledge about the form of $\sigma(v_r)$ can be employed.
It has been shown extensively \cite{liuGenerationWindowedMultipole2018, ducruWindowedMultipoleRepresentation2021} at this point that the cross section is accurately represented as a sum of poles
in addition to a low order Laurent expansion $(N\approx 7)$, vis:
\begin{equation}
  \sigma(E) = \frac{1}{E} \left(\Re \left[ \sum_j \frac{r_j}{p_j-\sqrt{E}} \right] + \sum_{n=0}^{N} a_n E^{n/2} \right)
  \label{eq:xs_form_full}
\end{equation}
In fact, for the purposes of sampling the resonance upscattering effect,
we claim and later numerically demonstrate that the narrow range of attainable $v_r$ leads the zero-kelvin cross section to be accurately represented
as a single pole and a linear term, over a sufficiently narrow range of energies:
\begin{equation}
  \sigma(E) = \frac{1}{E} \Re \left[  \frac{r_j}{p_j-\sqrt{E}} \right] + \sigma_0 + \sigma_1 \sqrt{E}
  \label{eq:xs_form}
\end{equation}
The $\sigma_0$ and $\sigma_1$ terms are calculated through a linearization process to avoid some complexities with higher order polynomial fitting.  An algorithm for finding the
best values of $\sigma_0$ and $\sigma_1$ is presented in section \ref{subsection:linearization}.

With this approximation, we use the technique developed in \cite{choiRelativeSpeedTabulation2021}:
rather than attempting to sample the target speed ($V$) and direction cosine ($\mu$), one instead
samples first the relative velocity $v_r$, and then samples $\mu$ from the distribution of $\mu$
conditioned on $v_r$. The distribution in this form for projectile speed $v$ is thus\cite{choiRelativeSpeedTabulation2021}:
\begin{equation}
  P(v_r|v) = \left(e^{-\beta^2(v-v_r)^2}-e^{-\beta^2(v+v_r)^2} \right)v_r^2 \sigma_0(v_r)
\end{equation}
Next, as previously shown for Doppler-broadening of the Windowed Multipole format \cite{joseyWindowedMultipoleCross2016} and the original full multipole format \cite{hwangRigorousPoleRepresentation1987},
we note the term $e^{-\beta^2(v+v_r)}$ to be negligible and use the following:
\begin{equation}
  P(v_r|v) \approx e^{-\beta^2(v-v_r)^2} v_r^2 \sigma_0(v_r)
  \label{eq:vr_distr}
\end{equation}


At this point, we write the multipole cross section in terms of the relative velocity rather than
in terms of energy. 
%
%
The formula to use is:
\begin{equation}
  \sigma(v_r) = \frac{2}{m_n v_r^2} \Re \left[  \frac{r_j}{p_j-\sqrt{2 m_n} v_r} \right] + \sigma_0 + \sigma_1 \sqrt{2m_n} v_r
  \label{eq:vel_xs}
\end{equation}
If we define the auxiliary variables $x=\beta(v_r-v)$ and $y=\beta v$, and insert Eq. \ref{eq:vel_xs}
in the marginal target collision rate distribution in terms of relative speed, Eq. \ref{eq:vr_distr},
some algebra reveals that:
\begin{equation}
  p(x|y) = e^{-x^2} \left(\Re\left[ \frac{\beta r_j}{z-x} \right] + \beta^{-2} (x+y)^2 \left(\sigma_0 + \sigma_1 x\right) \right)
  \label{eq:x_cond_distr}
\end{equation}
where $z = \beta p_j - y$, which represents a dimensionless measure of the energy gap between the center
of the Maxwell-Boltzmann distribution and the location of the resonance.

At this point, we can consider the CDF for the random variable $x$ (dimensionless
target velocity) conditioned on $y$ (dimensionless projectile velocity). Integrating Eq. \ref{eq:x_cond_distr} yields:
\begin{equation}
  C P(x|y) = \int_{-\infty}^x e^{-x'^2} \left(\Re\left[ \frac{\beta r_j}{z-x'} \right] + \beta^{-2} (x'+y)^2\left(\sigma_0 +\sigma_1 x'\right)\right) \dif x' 
\end{equation}
where $C$ is the normalizing constant. After distributing the integral
and interchanging the $\Re$ operator with integration, we obtain:
\begin{multline}
  C P(x|y) = \Re\left[ \frac{r_j \pi}{i \beta^{-1}} w(z, x) \right] + \frac{\beta^{-2} \sigma_0}{4}(-2 e^{-x^2} (x+2y) + \sqrt{\pi} (1+2y^2) (1+\erf(x))) + \\
\frac{1}{2\beta^2} \sigma_1 e^{-x^2} \left( 1+(x+y)^2 + \sqrt{\pi} y (1+\erf(x)) \right) 
  \label{eq:relative_speed_cdf}
\end{multline}
where the normalizing constant for the distribution is:
\begin{equation}
  C = \Re\left[\frac{r_j \pi}{i \beta^{-1}} w(z) \right] + \frac{\sqrt{\pi} }{2\beta^2} \left( \sigma_0 (1+2y^2) + \sigma_1 y \right) \quad,
  \label{eq:normalizing_constant}
\end{equation}
which we point out is nothing more than the Doppler-broadened scattering cross section
at temperature $T$ under the single pole approximation.


The new special function we deem the ``incomplete Faddeeva function'' is defined as:
\begin{equation}
  w(z, x) = \frac{i}{\pi} \int_{-\infty}^x \frac{e^{-t^2}}{z-t} \dif t
  \label{eq:inc_fad}
\end{equation}
And it can easily be seen that $w(z, \infty) = w(z)$ as per the definition of the Faddeeva function for $\Im[z]>0$:
\begin{equation}
  w(z) = \frac{i}{\pi} \int_{-\infty}^\infty \frac{e^{-t^2}}{z-t} \dif t
  \label{eq:faddeeva}
\end{equation}
Indeed, for this application, $\Im[z] > 0$ and we maintain this assumption going forward.
A specialized root finder has been developed to quickly invert this CDF and consequently sample the
relative velocity. This thus constitutes a method to sample target velocities without rejection sampling
or extensive tables.

The novelty in our approach lies entirely in the treatment of the zero kelvin cross section and
analytical representation of the relative speed cumulative distribution. After sampling from the
relative speed distribution, we must sample the target
polar angle distribution conditioned on the relative speed as done in \cite{choiRelativeSpeedTabulation2021}. For completeness, we conclude
with the CDF of the target speed:
\begin{equation}
  C(V|v_r) \propto \begin{cases}
    0 & V \leq |v_r - v| \\
    1-e^{-\beta^2 V^2} & |v_r-v| < V < v_r+v \\
    1  & v_r + v \leq V \\
  \end{cases} \,
\end{equation}
for which \cite{choiRelativeSpeedTabulation2021} provides a straightforward sampling technique.
The remaining work is purely numerical, particularly in requiring an efficient, reasonably accurate algorithm for the incomplete Faddeeva function $w(z, x)$.

\subsection{The Incomplete Faddeeva Function}
The forthcoming discussion explores the properties of the incomplete Faddeeva function,
with a particular focus on properties which can be leveraged to obtain efficient
numerical approximations to it.
We advise the reader that absorbing this section in depth is not necessary to grasp the claims and algorithms made herein.

The incomplete Faddeeva function, as defined by Eq. \ref{eq:inc_fad} is $w : \mathbb{C} \times \mathbb{R} \rightarrow \mathbb{C}$.
This section attempts to build some intuition as to how this function behaves as $z$ and $x$ individually vary.
In resonance upscatter treatment, $z$ parametrizes the location, height, and width of the resonance with respect to the Maxwell-Boltzmann distributed velocities.
Values of $\Re[z]=0$ correspond to scattering resonances exactly situated at the mean component of relative speed along the neutron's line of flight predicted by the Maxwell-Boltzmann
distribution. Values of $\Re[z]>0$ correspond to resonances at higher energies than the particle's energy, and therefore
induce preferential scattering with relative velocities higher than the incident velocity, and vice-versa for $\Re[z] < 0$. $x$ parametrizes
the dimensionless target relative speed. Small values of $\Im[z]$ imply tall, narrow 
resonances, with
a limit of $\Im[z]=0$ being a singularity representing a resonance of infinite cross section. Large values of $\Im[z]$ 
model wider, weaker resonances.

\begin{figure}
   \centering
   \begin{subfigure}[b]{0.45\textwidth}
       \centering
       \includegraphics[width=\textwidth]{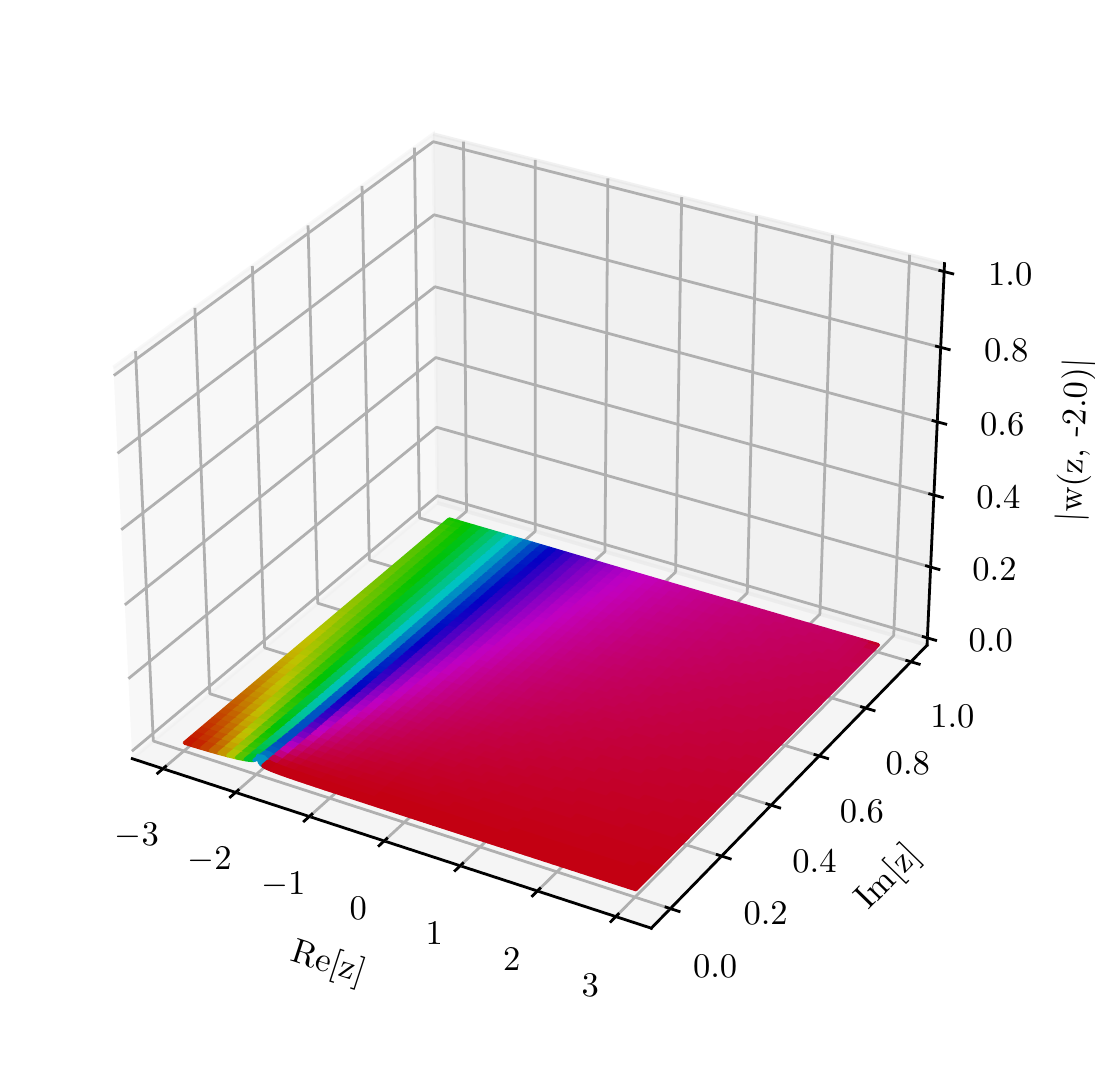}
       \caption{$w(z, -2.0)$}
   \end{subfigure}
   \begin{subfigure}[b]{0.45\textwidth}
       \centering
       \includegraphics[width=\textwidth]{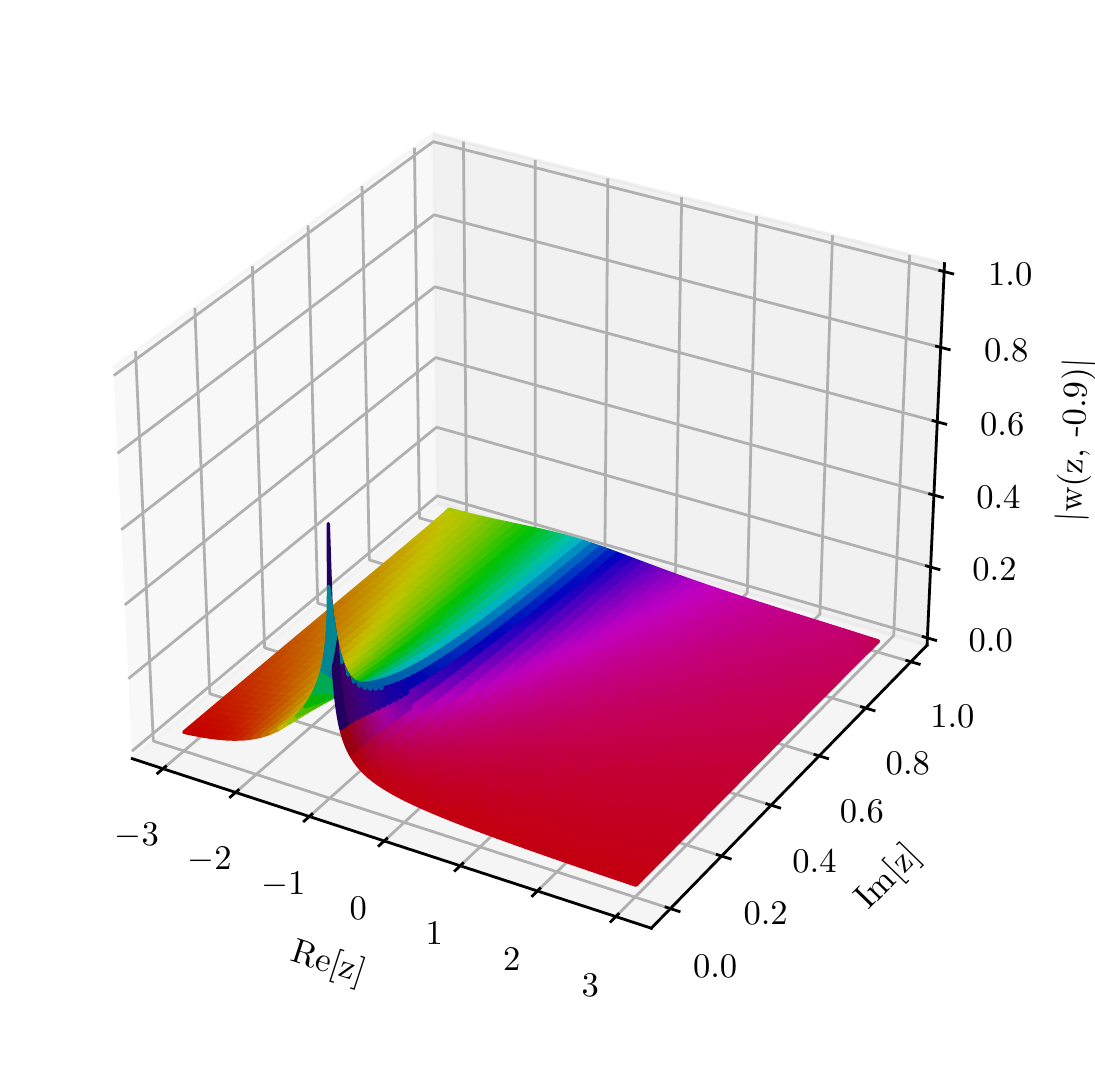}
       \caption{$w(z, -0.9)$}
   \end{subfigure}

   \begin{subfigure}[b]{0.45\textwidth}
       \centering
       \includegraphics[width=\textwidth]{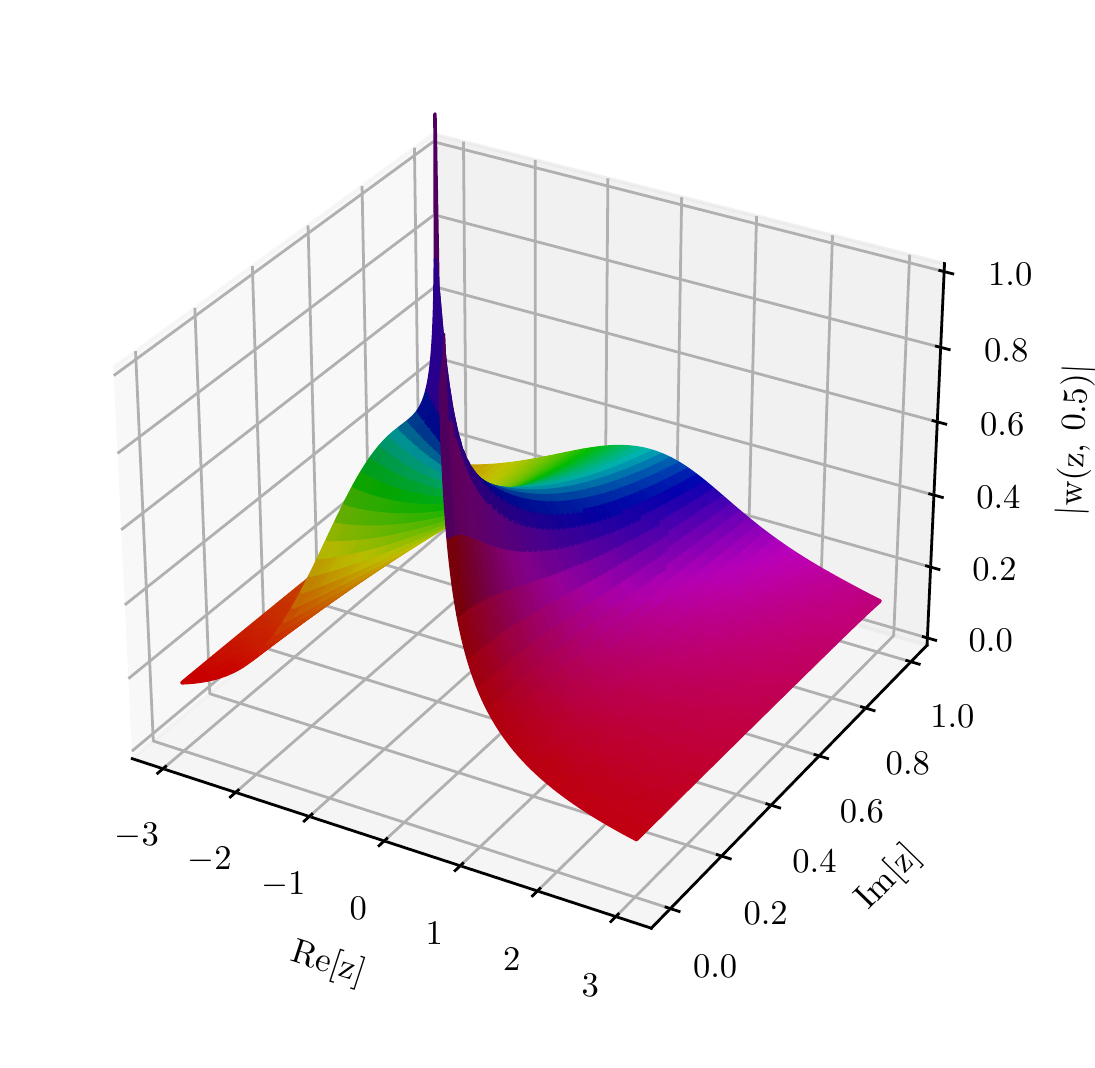}
       \caption{$w(z, 0.5)$}
   \end{subfigure}
   \begin{subfigure}[b]{0.45\textwidth}
       \centering
       \includegraphics[width=\textwidth]{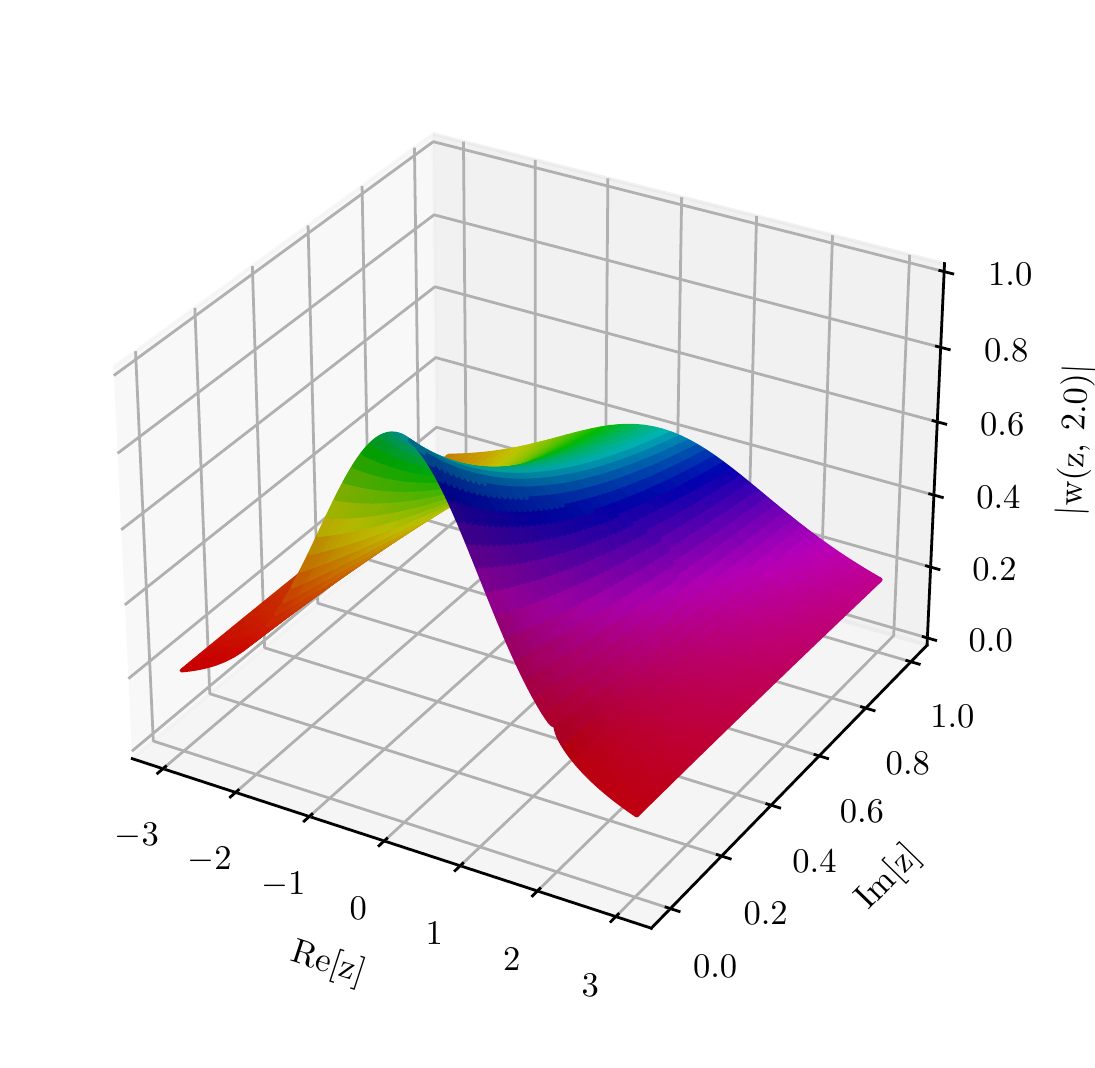}
       \caption{$w(z, 2.0)$}
   \end{subfigure}

   \caption{$w(z, x)$ for a few values of $x$. The height represents the magnitude, and coloring is done by phase.}
   \label{fig:inc_fad_plots}
\end{figure}

Fig. \ref{fig:inc_fad_plots} plots the incomplete Faddeeva function as a function of $z$ for a few values of $x$.
This illustrates the rapidly varying behavior of this function for small values of $\Im[z]$ when $\Re[z]\approx x$,
where a sharp peak follows the value of $x$ near the real line.
On top of that, it shows the approach to the familiarly shaped $w(z)$ as $x\rightarrow \infty$. The figure shows that $w(z, x)$ is
an increasing (in the sense of increasing in both real and imaginary part)
function in $x$ for many values of $z$, but not all. This behavior is explained by the following asymptotic analysis.

\begin{figure}
   \centering
   \begin{subfigure}[b]{0.45\textwidth}
       \centering
       \includegraphics[width=\textwidth]{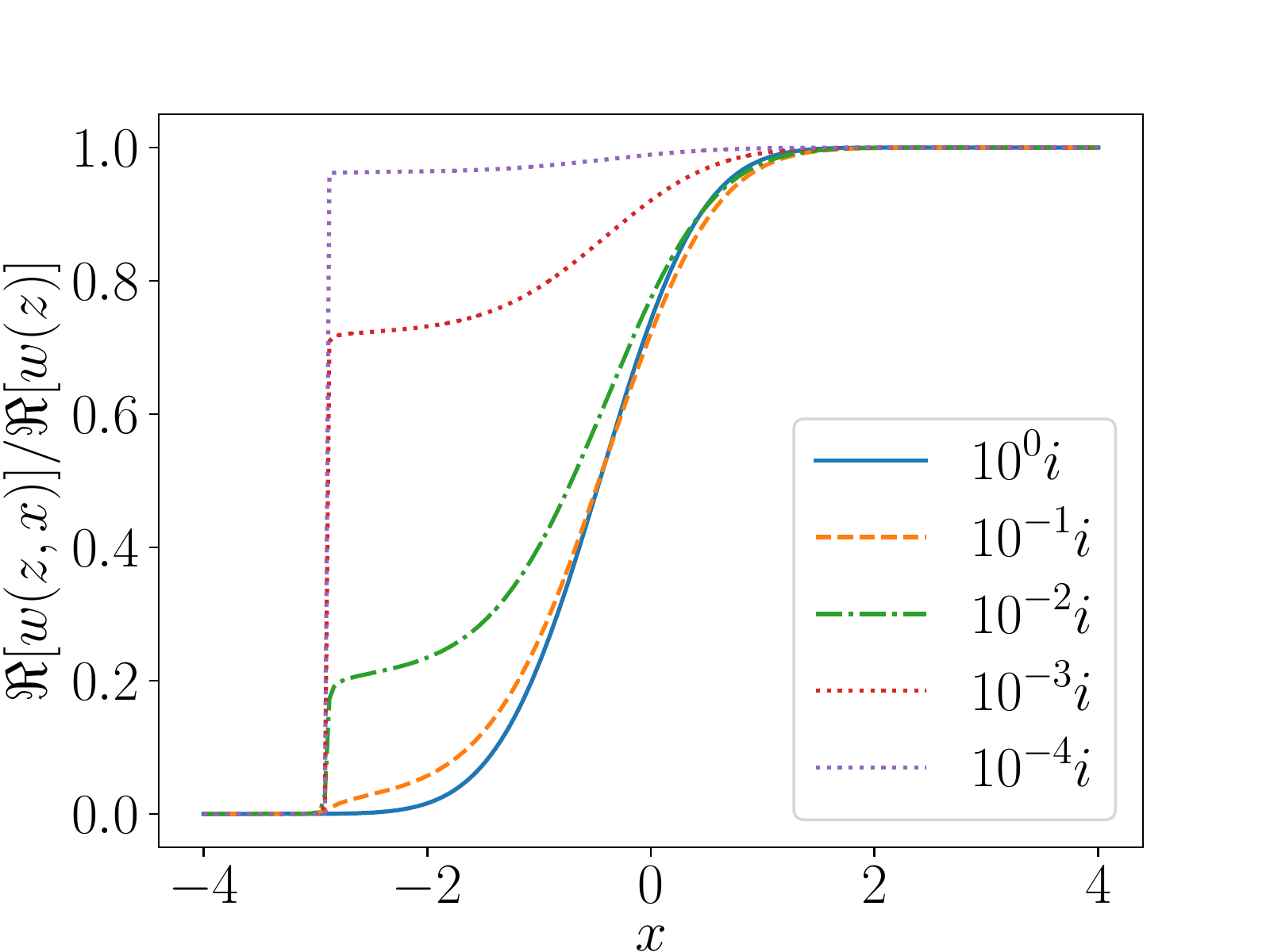}
       \caption{$\Re[z]=-2.9$}
   \end{subfigure}
   \begin{subfigure}[b]{0.45\textwidth}
       \centering
       \includegraphics[width=\textwidth]{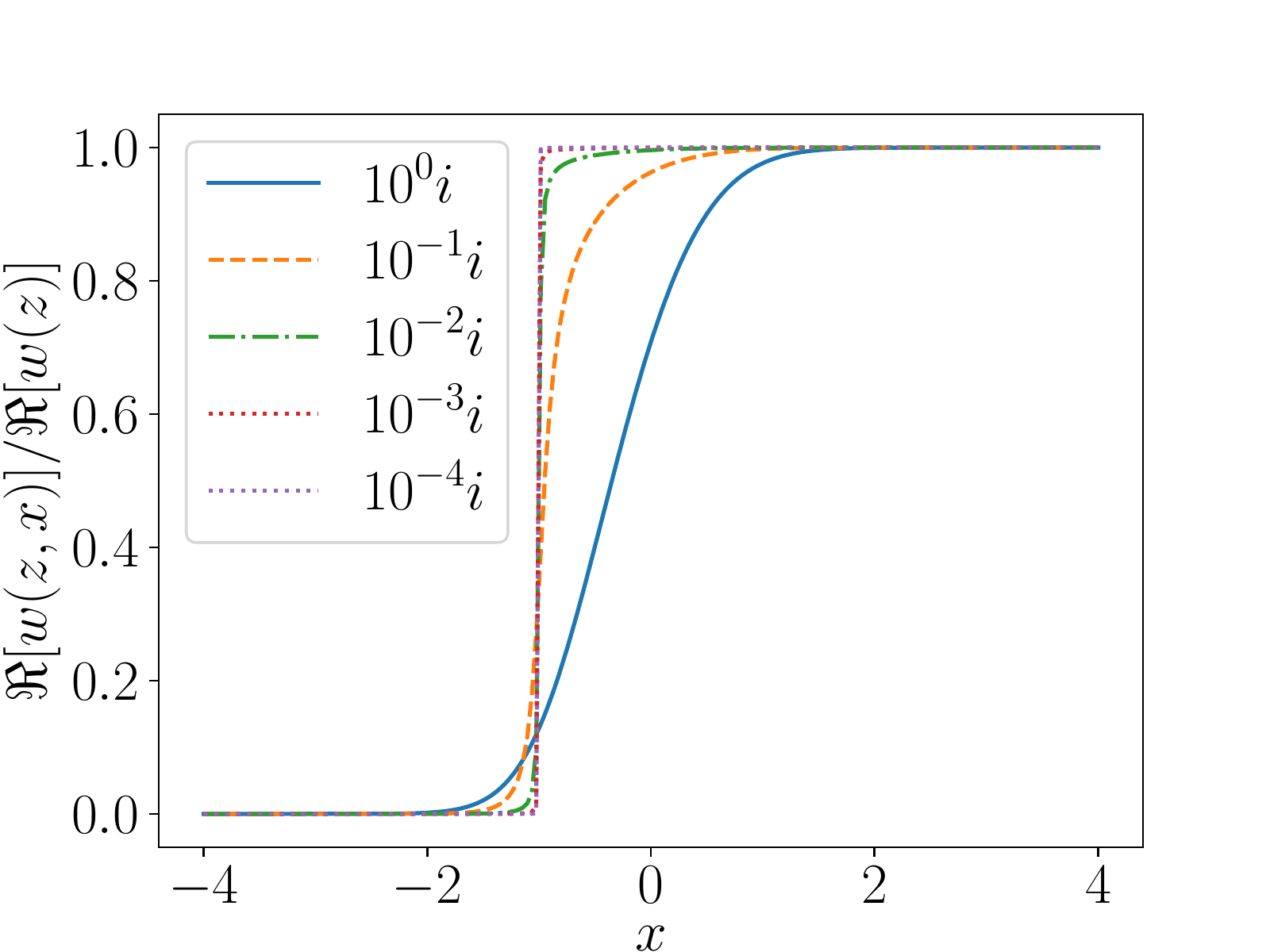}
       \caption{$\Re[z]=-1.0$}
   \end{subfigure}

   \begin{subfigure}[b]{0.45\textwidth}
       \centering
       \includegraphics[width=\textwidth]{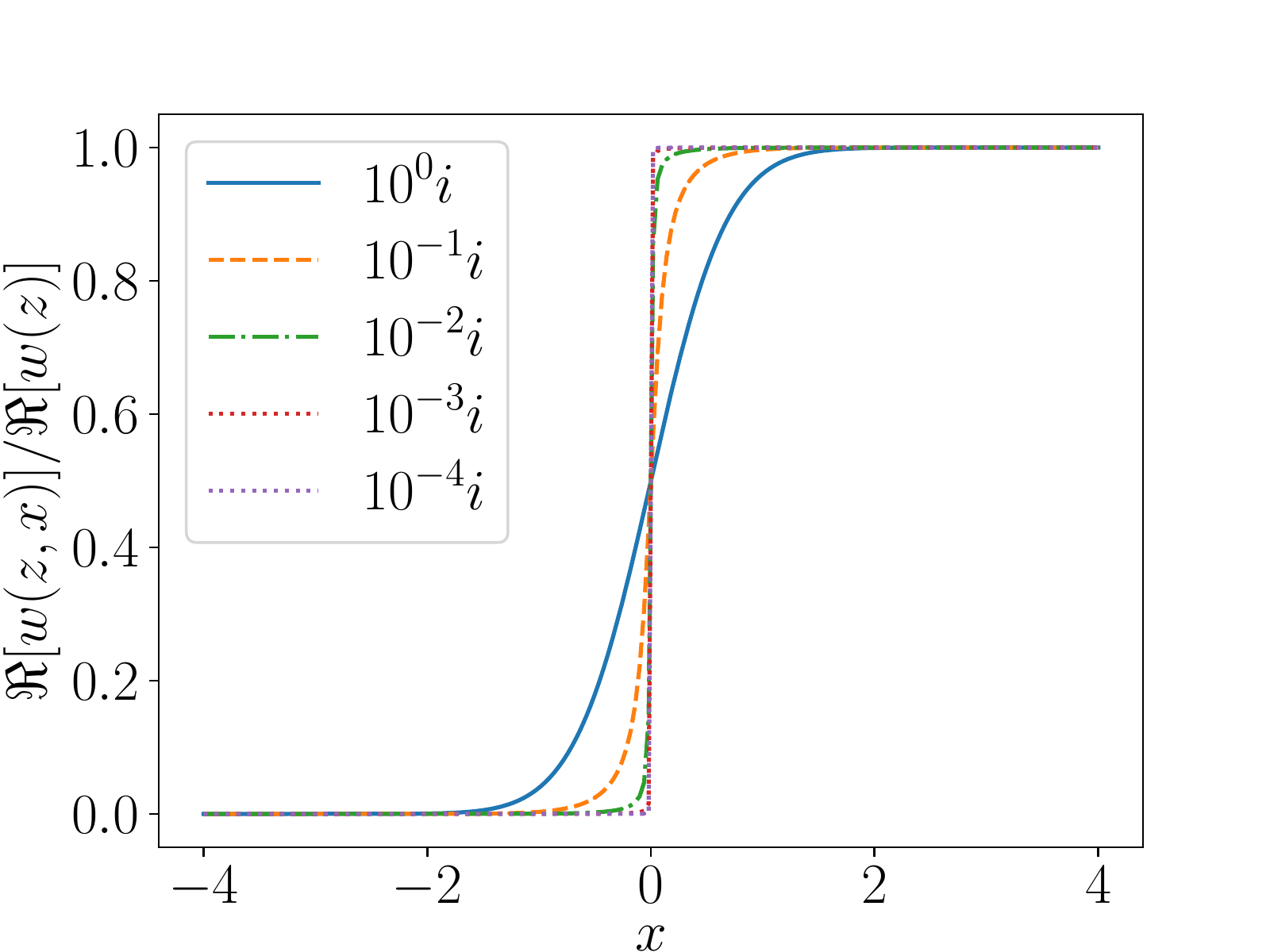}
       \caption{$\Re[z]=0.0$}
   \end{subfigure}
   \begin{subfigure}[b]{0.45\textwidth}
       \centering
       \includegraphics[width=\textwidth]{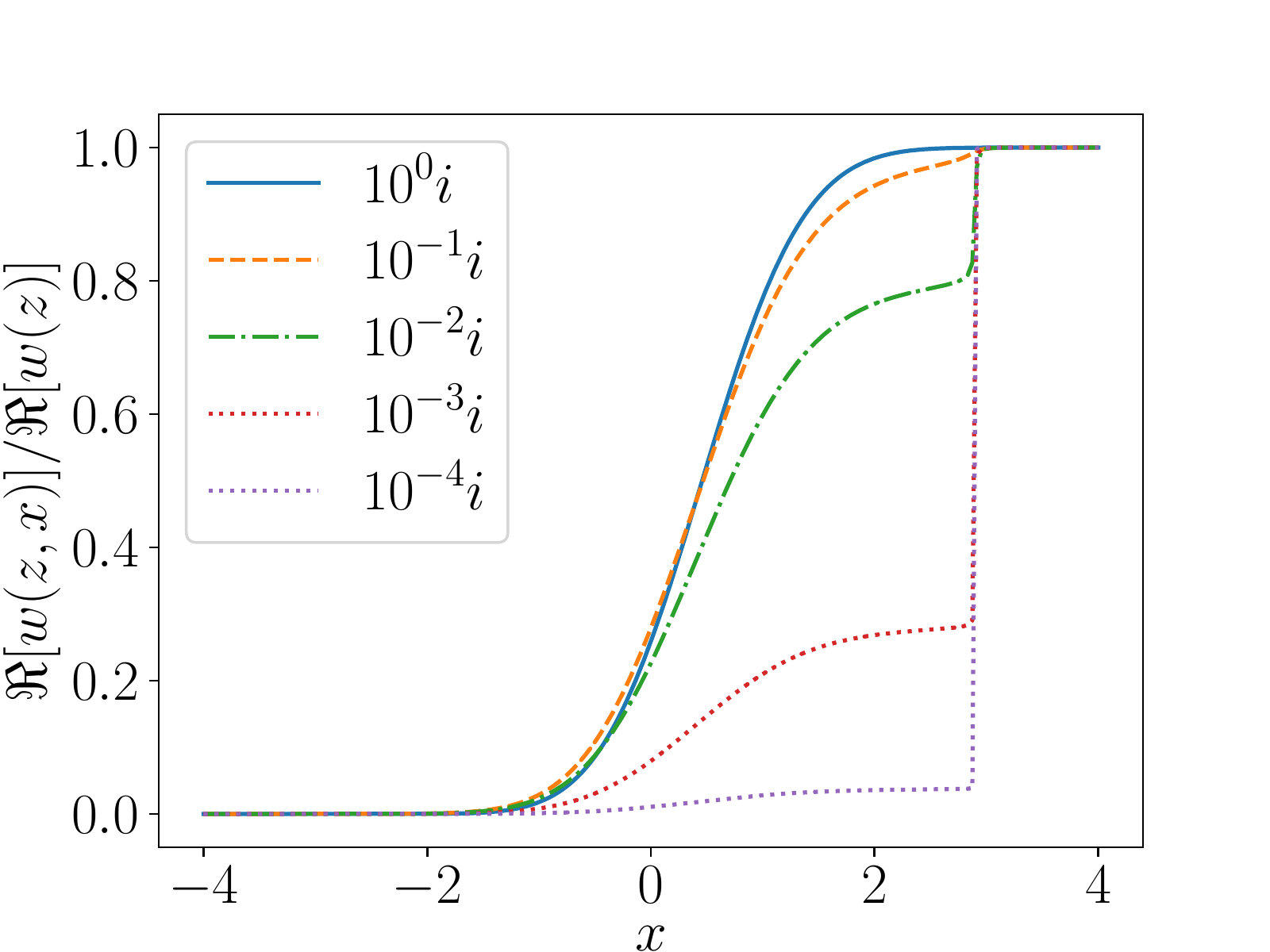}
       \caption{$\Re[z]=2.9$}
   \end{subfigure}

   \caption{$\Re[w(z, x)]$ for a few values of $z$. The legend is the imaginary number added to the real part specified in each figure's caption.
   The plotted quantity is normalized by $\Re[w(z)]$ so all lines tend to unity as $x$ grows.}
   \label{fig:inc_fad_rp_line_plots}
\end{figure}

Humli\^cek \cite{humlicekEfficientMethodEvaluation1979} provides the following asymptotic formula
for the Faddeeva function without derivation: 
\begin{equation}
  w(z) \approx \frac{i}{\sqrt{\pi} z} \quad.
  \label{eq:asymptotic_faddeeva}
\end{equation}
Considering the definition of the Faddeeva function once more in Eq. \ref{eq:faddeeva},
This approximation results from supposing that if $z$ is large, and that only values of $t$
near zero contribute to the integral, we can approximate $w(z)$ as:
\begin{equation}
  w(z) \approx \frac{i}{\pi} \int_{-\infty}^\infty \frac{e^{-t^2}}{z} \dif t
\end{equation}
which immediately yields the asymptotic estimate of Eq. \ref{eq:asymptotic_faddeeva}. Proceeding with the
same approximation in the context of the incomplete Faddeeva function, we thus obtain:
\begin{equation}
  w(z, x) \approx \frac{i}{2\sqrt{\pi} z} \left( \text{erf}(x)+1\right) \quad \quad \text{if} \quad |z| \gg 1
  \label{eq:inc_fad_apprx_1}
\end{equation}
Suggesting a close connection between the incomplete Faddeeva function's behavior in $x$ and the error
function. 
In fact, this hunch is confirmed by the following identity which connects the incomplete Faddeeva
to the standard Faddeeva function:
\begin{equation}
  w(z, x) = \frac{1}{2} (1+\erf(x)) w(z) + \frac{i e^{-x^2}}{\pi} \int_0^\infty \frac{e^{-t^2} e^{2i t z} \dif t}{i(x-z)+t}
  \label{eq:inc_fad_identity}
\end{equation}
Proof of this relation is provided in Appendix A. 

\begin{figure}
  \centering
  \includegraphics[width=0.6\textwidth]{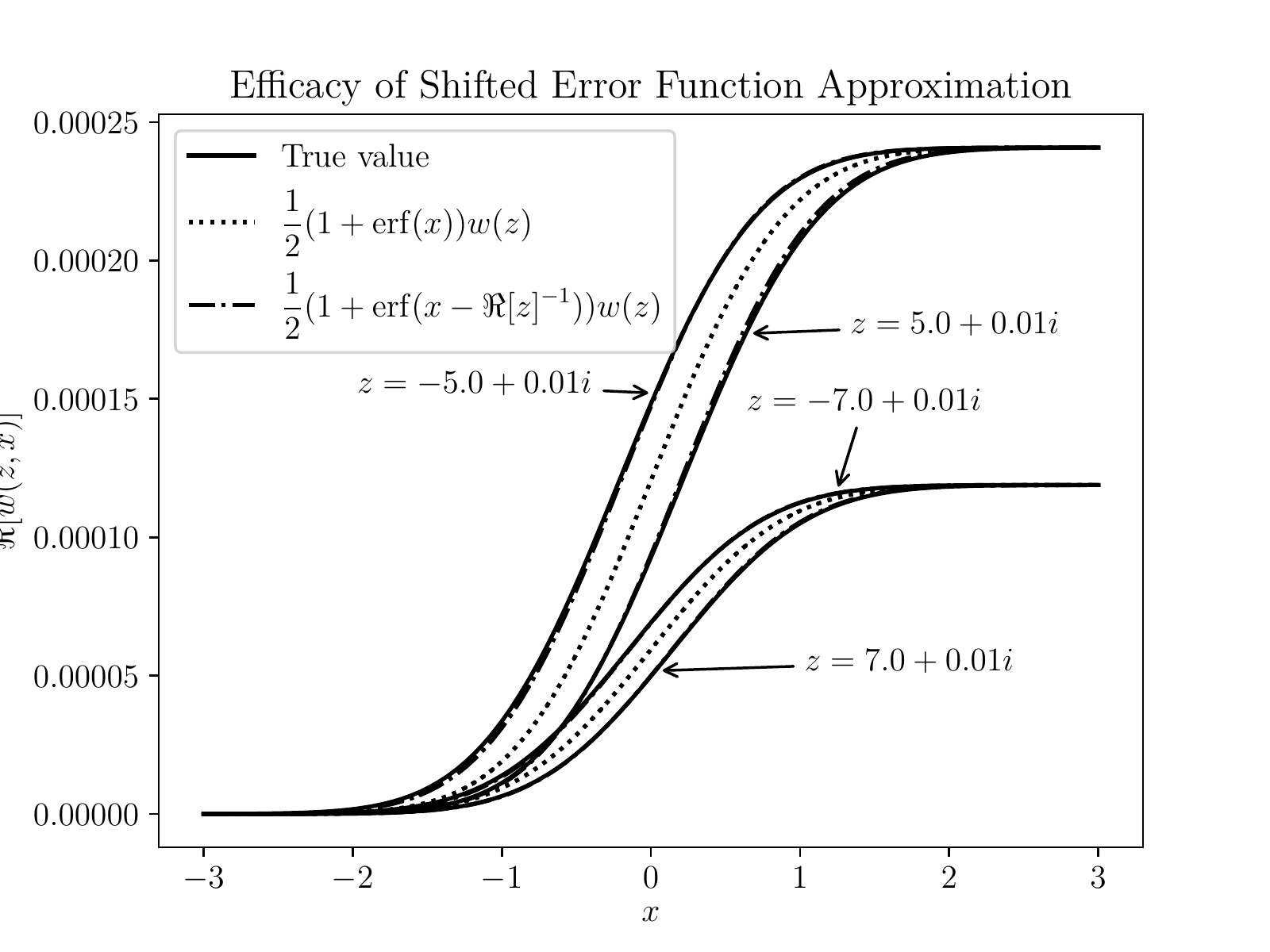}
  \caption{Accuracy of the $w(z, x)$ approximation for $|\Re[z]|>5$. An unshifted error function
  is shown for comparison, representing a more naive asymptotic approximation to $w(z, x)$.}
  \label{fig:asymp_inc_fad}
\end{figure}

An even more accurate asymptotic estimate can be obtained for large $\Re[z]$,
approximating the pole term as
\begin{equation}
  \frac{1}{z-t} \approx \frac{1}{2z} \left(1+e^{2t/z}\right) \quad.
\end{equation}
This matches the value, slope, and curvature with respect to $t$ of the pole about $t=0$.
Substituting this back to Eq. \ref{eq:inc_fad} and adjusting the expression such that $\Re[w(z, x)]$ is strictly
increasing (as suggested by Eq. \ref{eq:inc_fad_apprx_1}), and matching the asymptotic value of $w(z)$ as suggested
by Eq. \ref{eq:inc_fad_identity}, results in
\begin{equation}
  w(z, x) \approx \frac{1}{2} \left(1+\erf\left(x-\Re[z]^{-1}\right) \right) w(z) \quad.
  \label{eq:inc_fad_apprx_2}
\end{equation}
Eq. \ref{eq:inc_fad_apprx_2} is sufficiently accurate to be used in practical computations,
as shown by Fig. \ref{fig:asymp_inc_fad}.
In the context of resonance scattering, Eq. \ref{eq:inc_fad_apprx_2} shows that the relative speed
distribution gets shifted forward by a nondimensionalized factor of $1/\Re[z]$ with its influence
scaling by $\Re[r_j w(z)]$ times the resonances residue as suggested by Eq. \ref{eq:relative_speed_cdf}.

Another useful property of the incomplete Faddeeva function is a simple connection between its
derivative in the complex plane and its value. This is similar in nature to the derivative of the
Faddeeva function \cite{abramowitzHandbookMathematicalFunctions1965}:
\begin{equation}
  \od{w(z)}{z} = \frac{2i}{\sqrt{\pi}} - 2 z w(z)
  \label{eq:faddeeva_derivative}
\end{equation}
The relation we have obtained generalizes this as:
\begin{equation}
  \od{w(z, x)}{z} = \frac{i}{\sqrt{\pi}} \left(1+\erf(x)+\frac{e^{-x^2}}{\sqrt{\pi}(x-z)} \right)-2 z w(z, x)
  \label{eq:inc_faddeeva_derivative}
\end{equation}
which clearly maintains consistency with Eq. \ref{eq:faddeeva_derivative} as $x\rightarrow \infty$.
This fairly simple connection between the derivative of $w(z, x)$ and its value can be utilized
for efficient sensitivity analysis of the scattering kernel with respect to windowed multipole parameters.

The forthcoming discussion presents some further concepts in the direction of efficient numerical evaluation
of the incomplete Faddeeva function.
Going forward, we denote the second integral appearing in Eq. \ref{eq:inc_fad_identity} as:
\begin{equation}
  I(z, m) = \int_0^\infty \frac{e^{-t^2} e^{2i t z} \dif t}{im+t}
  \label{eq:i_integral}
\end{equation}
where $\mathbb{R} \ni m = x-z$.
While it may seem that a half-range Gauss-Hermite quadrature may work well to efficiently approximate Eq. \ref{eq:i_integral},
this is not the case. Firstly, complex exponentials would have to be calculated at each quadrature point. Secondly, as the
real part of $z$ grows, the integrand oscillates more. In practice, values of $\Re[z]>10$ are frequently
encountered, and low degree quadratures would not capture the oscillation. Additionally, given that the
value of $\Im[z]$ is small, the denominator becomes nearly singular when $x \approx \Re[z]$. In fact, when
$\Im[z] = 0$, $I(z, m)$ becomes a discontinuous function in $x$ when interpreted as a principal value integral.

Eq. \ref{eq:i_integral} is equivalent to the \textit{incomplete Goodwin-Staton integral}, referred to in \cite{deanoAnalyticalNumericalAspects2010}.
However, to our knowledge, only asymptotic analysis has been performed on this type of integral before, without
any development of numerical routines. Recent work in the field of finance \cite{aidStructuralRiskNeutral2013} presents results for computing
what the authors define as the \textit{extended incomplete Goodwin-Staton integral}, for which the $\nu=1$ case
is of interest in the present discussion. Unfortunately, the authors' numerical method works for all cases
except $\nu=1$, suggesting Eq. \ref{eq:i_integral} to be of a fundamentally different nature.

In our experience, the difficulty with large $\Re[z]$ cannot be ameliorated by a stationary phase technique
\cite{benderAdvancedMathematicalMethods2010}, as these tend to accentuate the pole behavior and remain of
similar difficulty for half range Gauss-Hermite quadrature.

\begin{figure}
  \centering
  \includegraphics[width=0.5\textwidth]{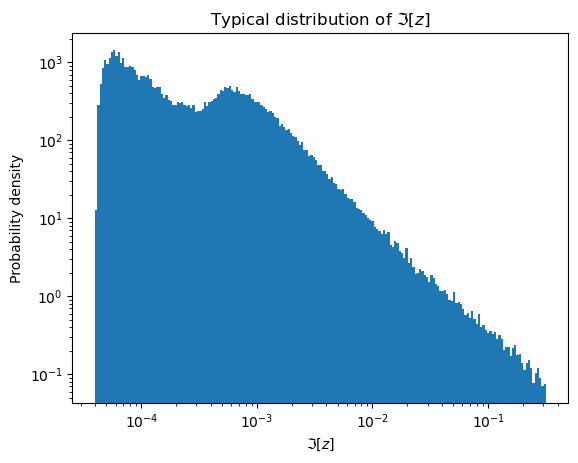}
  \caption{Distribution of imaginary part of the poles in the windowed multipole method. These were collected
  from every nuclide in OpenMC's regression testing dataset, based on ENDFVII.1.}
  \label{fig:imag_parts}
\end{figure}

In the windowed multipole method, $\Im[z]$ is near zero, as shown in Fig. \ref{fig:imag_parts}. Therefore, we
can expect to frequently encounter nearly singular integrands in Eq. \ref{eq:i_integral}.
An efficient numerical technique which explicitly treats this behavior can be devised by first noticing
that $I(z, m)$ satisfies this differential equation in the complex plane:
\begin{equation}
  \od{I}{z} + 2 z I = \frac{1}{x-z}
  \label{eq:i_integral_diffeq}
\end{equation}
This can be used to connect the value $I(z_0, x)$ at a point $z_0$ to another point $z_1$.
The integrating factor technique shows that:
\begin{equation}
  I(z_1, x) = \int_{z_0}^{z_1} \frac{e^{t^2-z_1^2} \dif t}{x-t} + e^{z_0^2-z_1^2} I(z_0, x) 
  \label{eq:integrating_factor_soln}
\end{equation}
If the distance between $z_0$ and $z_1$ is small, the exponential in the numerator of Eq.
\ref{eq:integrating_factor_soln} can be Taylor expanded about $z_0$ to yield an efficient numerical scheme.

We have found that the imaginary part of $I(\Re[z], x)$ can be calculated
with a closed-form, discontinuous-in $x$ formula. Because the nearly discontinuous behavior of $I(z, x)$ in $x$
is the main source of difficulty here, Eq. \ref{eq:integrating_factor_soln} can
be used to resolve this behavior accurately after calculating the value of $I(\Re[z], x)$.
The real part of $I(\Re[z], x)$ is continuous in $x$ but not available in formula in terms of elementary functions; however, it is readily amenable to numerical approximation. In conclusion,
the real and imaginary part of $I(\Re[z], x)$ contrast each other: the former is readily
numerically approximated by series or similar methods, whereas the latter is discontinuous
and therefore not amenable to series or rational function approximation, and fortunately
has an exact formula.

The first goal at hand is to calculate $\Im[I(z, m)]$ for real values of $z$.
This can be written as:
\begin{equation}
  \Im[I(z, m)] = \int_0^\infty \frac{e^{-t^2}\left( t \sin(2z t) - m \cos(2 z t) \right) \dif t}{t^2+m^2} \quad.
\end{equation}
The linearity of integration can be distributed over both trigonometric terms. Each of the resulting integrals
can be found as respective sine and cosine transform integrals, which are listed in  \cite{f.g.;harrybatemanerdelyiTablesIntegralTransforms1954}.
Combining results from the sine and cosine transform tables yields:
\begin{equation}
  \Im[I(z, m)] = \pi e^{-z^2 + (m+z)^2} \left(\frac{1}{2} \erf(m+z) - \text{sign}(m)\right) \quad z\in \mathbb{R}
  \label{eq:i_imag_part}
\end{equation}

Next, we can find an approximation for the real part of $I(z, m)$:
\begin{equation}
  \Re[I(z, m)] = \int_0^\infty \frac{e^{-t^2}\left( t \cos(2z t) + m \sin(2 z t) \right) \dif t}{t^2+m^2} \quad.
\end{equation}
This expression can neither be written terms of elementary nor special functions, to our knowledge.
In order to manipulate it to obtain a numerical expression, consider the auxiliary function:
\begin{equation}
  R(z, m) = \int_0^\infty \frac{e^{-t^2} \sin(2 z t) \dif t}{t^2 + m^2}
  \label{eq:r_integral}
\end{equation}
which again, of course, cannot be represented in terms of elementary or special functions. This pinpoints the difficulty in calculating the real part of $I(z, m)$ because:
\begin{equation}
  \Re[I(z, m)] = m R(z, m) + \frac{1}{2} \pd{R}{z} \quad.
\end{equation}
We thus seek a straightforwardly differentiable approximation to $R(z, m)$. The intuition behind our forthcoming numerical approximation to $R(m, z)$
comes from the fact that for $m\gg 1$, $t$ is negligible in the denominator compared to $m$ over the range where
the $e^{-t^2}$ weighting is large, hence
\begin{equation}
  R(m, z) \approx \frac{1}{m^2}\int_0^\infty e^{-t^2} \sin(2 z t) \dif t = \frac{1}{m^2} F(z)
\end{equation}
where $F(z)$ is the Dawson F function \cite{abramowitzHandbookMathematicalFunctions1965}. By standard
asymptotic analysis, matching the $z$ derivative at $z=0$ and the $z\gg 1$ asymptote for the Dawson F function
results in the improved estimate:
\begin{equation}
  R(m, z) \approx \frac{1}{m} \sqrt{e^{m^2} E_1(m^2)} F\left( m z \sqrt{e^{m^2} E_1(m^2)} \right)
\end{equation}
which is accurate to within about 5\% across the full range of $m$ values. However, this level
of accuracy is not appropriate for engineering calculations. 


To build more intuition for $R(m, z)$, its close
relation to the Dawson F function is evinced by considering the Maclaurin series in $z$ for both:
\begin{equation}
  F(z) = z - \frac{2}{3} z^3 + \frac{4}{15} z^5 - \frac{8}{105} z^7 + \cdots \quad,
\end{equation}
\begin{equation}
  R(m, z) = e^{m^2}\left(E_1(m^2) z - \frac{2}{3}E_2(m^2) z^3 + \frac{4}{15}E_3(m^2) z^5 - \frac{8}{105}E_4(m^2) z^7 + \cdots \right)
\end{equation}
which further highlights the utility of maintaining consistency of $R(m, z)$ with its asymptotic sister $F(z)$.
The series are the same, but with the addition of exponential integral multiplying factors in $R(m, z)$.

In order to find such a numerical relation which maintains this consistency in the asymptotic case,
Eq. \ref{eq:r_integral} can be transformed via interchange of differentiation and integration.
Consider the generalized function:
\begin{equation}
  R(m, z, \alpha) = e^{m^2} \int_0^\infty \frac{e^{-\alpha(t^2+m^2)} \sin(2 z t) \dif t}{t^2+m^2}
\end{equation}
Differentiating reveals that:
\begin{equation}
  \pd{R(m, z, \alpha)}{\alpha} = -e^{m^2}\int_0^\infty e^{-\alpha (t^2+m^2)} \sin(2 z t) \dif t = -\frac{ e^{-m^2(\alpha-1)}}{\sqrt{\alpha}}  F\left(\frac{z}{\sqrt{\alpha}}\right)
\end{equation}
The fundamental theorem of calculus then applies:
\begin{equation}
  R(m, z, \infty) - R(m, z, 1) = e^{m^2} \int_1^\infty \frac{1}{\sqrt{\alpha}} e^{-m^2 \alpha} F\left(\frac{z}{\sqrt{\alpha}}\right) \dif \alpha
  \label{eq:r_integral_v2}
\end{equation}
Using the fact that $R(m, z, \infty) = 0$ and doing a change of variables, Eq. \ref{eq:r_integral} becomes:
\begin{equation}
  R(m, z) = 2 e^{m^2} \int_1^\infty e^{-m^2 t^2} F(z/t) \dif t
  \label{eq:r_integral_v3}
\end{equation}
which confirms our hunch about the close relation of $F(z)$ to $R(m, z)$; it is an infinite superposition of
stretched and scaled Dawson F functions.


We have found success in inserting an approximation for 
$F(z)$ to Eq. \ref{eq:r_integral_v3}. Well-known approximations to $F(z)$ based on rational expressions
and other elementary functions \cite{letherConstrainedNearminimaxRational1997, letherElementaryApproximationsDawson1991}
do not result in numerically useful expressions. However, noting that:
\begin{equation}
  \int_1^\infty e^{-m^2 t^2} (z/t)^{(2k+1)} \dif t = \frac{1}{2} z^{1+2k} E_{1+k}(m^2)
\end{equation}
it can be seen that approximations to $F(z)$ in the form a power series can be computationally efficient
in light of the recursion relation for exponential integrals:
\begin{equation}
  n e^{m^2} E_{n+1}(m^2) = (1 - m^2 e^{m^2} E_n(m^2))
  \label{eq:expint_recurrence}
\end{equation}
Careful attention must be paid to the floating point properties of this relation \cite{gautschiRecursiveComputationCertain1961}.
The magnification of computational errors grows arbitrarily large, and we later present a specialized numerical algorithm
guaranteeing floating point stability.

In order to thus obtain a simple, efficient approximation to $R(m, z)$, we employ the Chebyshev expansion
valid for $z\in[-5, 5]$ presented in \cite{hummerExapansionsDawsonFunction1964}. The results of our method
could be improved by using a finer piecewise division for the Chebyshev expansion of $F(z)$ as in \cite{codyChebyshevApproximationsDawson1970},
but we have used the present approach for simplicity of implementation. Thus, if $F(z)$'s truncated Chebyshev expansion is converted
to the power series basis:
\begin{equation}
  F(z) \approx \sum_{i=0}^n c_n (z/5)^{2*i+1}
  \label{eq:fz_apprx}
\end{equation}
we obtain approximations of the form
\begin{equation}
  R(m, z) \approx \sum_{i=0}^n \frac{c_n}{2}E_{1+i}(m^2) (z/5)^{2*i+1} \quad.
  \label{eq:r_approximation}
\end{equation}


\subsection{Numerical Implementation}
\subsubsection{Stable, Efficient Calculation of an $E_n$ Sequence}
The magnification of error in the forward recurrence relation for exponential integrals from \cite{gautschiRecursiveComputationCertain1961}
is:
\begin{equation}
  |\rho_n| = \frac{x^n E_1(x)}{n! E_{n+1}(x)}
\end{equation}
Notably, the error magnification of the reverse recurrence relation is the reciprocal of this quantity. Moreover, $|\rho_n|$ is a function increasing from 1, reaching a maximum, and monotonically descending below 1 \cite{gautschiRecursiveComputationCertain1961}.
As a consequence, a critical index $n^*$ exists such that iterating outward from it results in a numerically
stable recursion algorithm. In terms of evaluating Eq. \ref{eq:r_approximation}, this means splitting the polynomial
in $z$ into parts above the index $n^*$ and those below. After computing $e^{m^2} E_{n^*}(m^2)$, Horner's method is used in the reverse recurring relation
down to the term of order $z$, and forward recursion is employed to evaluate the polynomial of degree leading from $2n^*+1$
up to $2n+1$. 

\begin{figure}
  \centering
  \includegraphics[width=0.5\textwidth]{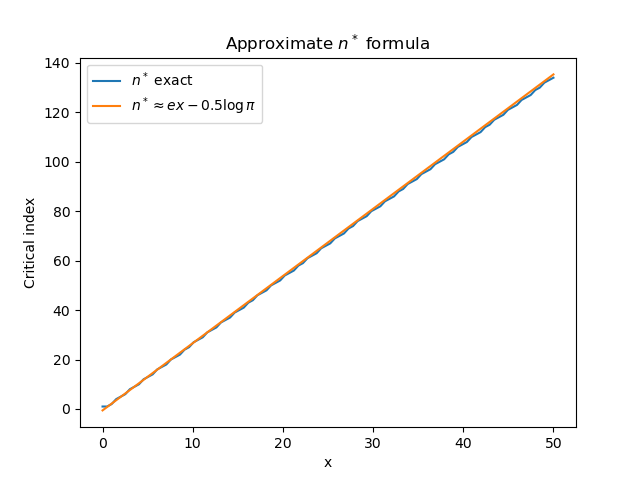}
  \caption{Approximate solution to finding the first value of $n$ such that
  $\frac{x^{n} E_1(x)}{n! E_{n+1}(x)} < 1$.}
  \label{fig:root_apprx}
\end{figure}

A simple result we have obtained is that the smallest value of $n^*$ such that:
\begin{equation}
  \frac{x^{n^*} E_1(x)}{n^*! E_{n^*+1}(x)} < 1
\end{equation}
is well approximated by:
\begin{equation}
  n^* \approx e x - \frac{1}{2} \log\pi \quad.
  \label{eq:asymp_nstar}
\end{equation}
Appendix \ref{appendix:asymp_nstar} shows how this can be obtained.
Fig. \ref{fig:root_apprx} illustrates the accuracy of Eq. \ref{eq:asymp_nstar}.
Using this information,
algorithm \ref{alg:r_integral} explicitly states the procedure to calculate $R(m, z)$ and its derivative with respect to $z$.
While the use of a power basis polynomial is sub-optimal, numerically speaking, the main source of numerical error in this
scenario originates from the recursive exponential integral formula. The specification of the algorithm assumes that
an accurate method for computing $E_n(x) e^x$ has been provided, which is well documented in many other works.
We have employed a C++ adaptation of the continued fraction approximation employed by the Cephes library \cite{Cephes},
which is documented in \cite{abramowitzHandbookMathematicalFunctions1965}.

\subsubsection{The Jump Integral}
After computing the value of $I(\Re[z], m)$, the differential equation of Eq.
\ref{eq:i_integral_diffeq} which $I$ follows can be used to calculate $I(z,
m)$. We calculate $I(z, m)$ in this manner due to the nearly discontinuous
behavior of $I(z, m)$; it has a jump discontinuity about $m=0$ if $z\in
\mathbb{R}$. Because the imaginary part of $z$ is small in windowed multipole
libraries, the resulting behavior is nearly discontinuous and hence is not
captured efficiently by general approximation techniques; finely
resolved tables or high polynomial orders would be required. Our approach thus
resolves the discontinuous component exactly with the piecewise function Eq.
\ref{eq:i_imag_part}. The nontrivial part of Eq. \ref{eq:integrating_factor_soln}
is the transcendental integral:
\begin{equation}
  J(z, x) = e^{-\Re[z]^2} \int_{\Re[z]}^z \frac{e^{t^2} \dif t}{x-t}
  \label{eq:jump_integral}
\end{equation}
We have deemed this term the \textit{jump integral} because it allows jumping
from values of $I(z, m)$ on the real line to values above the real line in the complex plane.
While it seems that our issue of approximating the transcendental integral $w(z, x)$ has seemingly not been
heretofore ameliorated due to the appearance of yet another transcendental integral Eq. \ref{eq:jump_integral},
a change of variables puts it into a form suitable for numerical approximation:
\begin{equation}
  J(z, x) = \int_0^{i \Im[z]} \frac{e^{u^2 + 2 u \Re[z]} \dif u}{m - u}
  \label{eq:jump_integral2}
\end{equation}
where again, $m=\Re[z]-x$. Because $\Im[z]$ is small as shown by Fig. \ref{fig:imag_parts}, the argument
to the exponential term is similarly small. Where this integral is well-defined ($m\neq 0$), the exponential term
can be expanded in its Maclaurin series and integrated term by term:
\begin{equation}
  e^{u^2 + 2 u \Re[z]} = \sum_{n=0}^\infty \frac{a_n}{n!} u^n \quad.
\end{equation}
It is verified that the coefficients $a_n$ satisfy the two-term recurrence:
\begin{equation}
  a_{n+1} = 2 \Re[z] a_n + 2(n-1) a_{n-1}; \quad a_0 = 1; \quad a_1 = 2 \Re[z]
\end{equation}
Next, the term-by-term integrals appear in the form: 
\begin{equation}
  \int_0^{i \Im[z]} \frac{u^n \dif u}{m-u} = m^n B_{i \Im[z] / m}(1+n, 0)
\end{equation}
Where $B_x(\cdot, \cdot)$ is the incomplete beta function,
defined as:
\begin{equation}
  B_x(a, b) = \int_0^z t^{a-1}(1-t)^{b-1} \dif t \quad.
\end{equation}
A recursion formula derived as a special case of formulas in
\cite{abramowitzHandbookMathematicalFunctions1965} efficiently calculates these
incomplete beta function values of higher $n$ in sequence:
\begin{equation}
  B_x(n+1, 0) = B_x(n, 0) - \frac{x^n}{n}
\end{equation}
In combination with the fact that:
\begin{equation}
  B_x(1, 0) = -\log(1-x) \quad,
\end{equation}
this yields an efficient numerical scheme for evaluating an integral of the truncated Maclaurin
series of the exponential of Eq. \ref{eq:jump_integral2}. Algorithm \ref{alg:jump_integral}
details the combination of all of these facts for an efficient approximation to $J(m, z)$.
This approximation works very well for problems with $|\Re[z]|\leq 5$, which easily covers the range
of scattering events where resonances appreciably affect the double differential at temperature.
Outside of that range, the integral becomes increasingly oscillatory, so an asymptotic approximation
is employed for $|z|>5$. This approximation is documented in Appendix \ref{appendix:jump_asymptotic}.

\begin{algorithm}
  \caption{Stable $R(m, z)$ approximation for $|z| \leq 5$. Appendix \ref{appendix:r_integral_asymptotic}
    is used for $|z| > 5$.}
  \label{alg:r_integral}
  \SetAlgoLined
  \SetKwInOut{Input}{Input}\SetKwInOut{Output}{Output}
  
  \Input{$m \in \mathbb{R}$, $z \in \mathbb{R}$}
  \Output{$R(m, z)$ from Eq. \ref{eq:r_integral} and $\pd{R}{z}$}
  \KwData{$c_n$ coefficients of Eq. \ref{eq:fz_apprx}, $1\leq n \leq n_{max}$}
  $n^* \leftarrow \min(\max(e m^2 - 0.57, 1), n_{max}) $\;

  expnexp $\leftarrow E_{n^*}(m^2) e^{m^2}$\;
  expnexp\_orig $\leftarrow $ expnexp\;
  result $\leftarrow 0$\;
  derivative $\leftarrow 0$\;
  
  \tcp{Backward recurse with Horner scheme}
  \For{$n\leftarrow n^*-1$ \KwTo $1$}{
    expnexp $\leftarrow \left(1-n \text{expnexp}\right)/m^2$\;
    derivative $\leftarrow \text{derivative} \cdot \text{pow}(z/5,2) + $ result\;
    result $\leftarrow (\text{result}) \text{pow}(z/5,2) +  \text{expnexp} \cdot c_{n-1}$ \;
  }
  derivative $\leftarrow 2 \text{derivative} \cdot \text{pow}(z/5,2)$ \;
  derivative $\leftarrow \text{derivative} + \text{result}$ \;
  result $\leftarrow \text{result} \cdot z/5$ \;

  \tcp{Forward recursing polynomial evaluation}
  x2 $\leftarrow \text{pow}(z/5, 2n^*-1)$\;
  x1 $\leftarrow \text{pow}(z/5, 2(n^*-1))$\;
  expnexp $\leftarrow$ expnexp\_orig\;
  \For{$n\leftarrow n^*$ \KwTo $n_{max}$}{
    result $\leftarrow c_{n-1} \cdot $expnexp $\cdot$ x2\;
    derivative $\leftarrow c_{n-1} \cdot $expnexp $\cdot$ x1 $\cdot (2n-1)$\;
    x2 $\leftarrow x2 \cdot \text{pow}(z/5, 2)$\;
    x1 $\leftarrow x1 \cdot \text{pow}(z/5, 2)$\;
    expnexp $\leftarrow (1-m^2 \cdot \text{expnexp})/n$\;
  }
  return (result, derivative)\;
\end{algorithm}

\begin{algorithm}
  \caption{Efficient $J(m, z)$ Approximation for $\Im[z]<1$ and $|\Re[z]|<5$. Appendix \ref{appendix:jump_asymptotic}
  describes the approximation for $|\Re[z]|>5$.}
  \label{alg:jump_integral}
  \SetAlgoLined
  \SetKwInOut{Input}{Input}\SetKwInOut{Output}{Output}
  
  \Input{$x \in \mathbb{R}$, $z \in \mathbb{C}$}
  \Output{$J(m, z)$ from Eq. \ref{eq:jump_integral}}

  result $\leftarrow 0$\;
  a0 $\leftarrow 1.0$\;
  a1 $\leftarrow 2 \Re[z]$\;

  result $\leftarrow $ b1 $\cdot$ a0\;
  b1 $\leftarrow b1 -i \Im[z]/(x-\Re[x])$\;

  \tcp{$n_{max}$ adjusts the number of truncated series terms}
  \For{$n\leftarrow 2$ \KwTo $n_{max}$}{
    tmp $\leftarrow 2 \Re[z] \cdot \text{a1} + 2 (n-1) \cdot \text{a0}$\;
    a0 $\leftarrow$ a1\;
    a1 $\leftarrow$ tmp\;
    result $\leftarrow \text{result} + \text{b1} \cdot \text{a1} \frac{m^n}{n!}$\;
    b1 $\leftarrow \text{b1} - \frac{1}{n+1} \left(\frac{i \Im[z]}{(x-\Re[x])}\right)^n$\;
  }

  return result\;
\end{algorithm}

\begin{algorithm}
  \caption{Efficient $w(z, x)$ approximation for $\Im[z]<<1$.}
  \label{alg:inc_fad_alg}
  \SetAlgoLined
  \SetKwInOut{Input}{Input}\SetKwInOut{Output}{Output}
  
  \Input{$z \in \mathbb{C}$, $x \in \mathbb{R}$}
  \Output{$w(z, x)$ from Eq. \ref{eq:inc_fad}}

  m $\leftarrow x - \Re[z]$\;
  \tcp{$R(m, z)$ integral and its derivative}
  rmz, drmzdz $\leftarrow$ call(Alg. \ref{alg:r_integral})\;
  \tcp{$I(m, z)$ integral}
  imz $\leftarrow m \cdot \text{rmz} + \frac{1}{2} \text{drmzdz} + \frac{i \pi}{2} e^{-\Re[z]^2+x^2} \left( \text{erf}(x)-\text{sign}(m)\right)$\;
  \tcp{$J(m, z)$ integral}
  ji $\leftarrow $ call(Alg. \ref{alg:jump_integral})\;
  imz $\leftarrow $ imz+ ji\;
  imz $\leftarrow $ imz$\cdot e^{\Re[z]^2-z^2}$\;
  result $\leftarrow \frac{i}{\pi} e^{-x^2} \cdot \text{imz}$\;
  result $\leftarrow \text{result} + \frac{1}{2} \left(\text{erf}(x)+1\right) w(z)$\;
  return result;
\end{algorithm}

Lastly, Algorithm \ref{alg:inc_fad_alg} gives the overall algorithm to compute $w(z, x)$
efficiently. It relies on access to some implementation of calculating $w(z)$, e.g.
the permissively licensed \cite{johnsonFaddeevaPackage} which implements a variety of approximations
to achieve high accuracy, or one of the various rational
approximations \cite{schreierVoigtComplexError2018, abrarovEfficientAlgorithmicImplementation2011,
humlicekEfficientMethodEvaluation1979} when higher error is permitted.
Regardless of the chosen $w(z)$ implementation,
our algorithm maintains asymptotic consistency such that $\lim_{x\rightarrow \infty} w(z, x) = w(z)$.  This work leverages a recent approximation tailored for WMP
\cite{forgetPerformanceImprovementsWindowed2022}.

\subsection{The Pole Sampling Approximation}
A key approximation of our technique that enables its computational efficiency is viewing the multipole
cross section in the relative speed distribution as a mixture distribution. The theoretical justification
is that if poles are present and sufficiently close to the incident neutron energy ($|z|<20$ specifically),
the relative speed PDF is well-approximated by ignoring the polynomial contribution: 
\begin{equation}
  P(x|y) \approx e^{-x^2} \Re\left[\sum_{j\in W(\beta^{-2} y^2)} \frac{\beta r_j}{z_j-x}\right]
\end{equation}
This expression is not employed to actually sample the scattering distribution. Rather, it is best
viewed as a mixture distribution in which each pole contributes a probability proportional to:
\begin{equation}
  \mathbb{P}[\sigma_s(x) = \frac{\beta r_j}{z-x}+\sigma_{0,j}+\sigma_{1,j}x] \propto \Re \left[ r_{j} w(z_j) \right]
\end{equation}
which defines a discrete distribution. In order to avoid the need for auxiliary storage and the calculation
of a normalizing constant to this distribution, we recommend finding the maximum of $-|r_j w(z_j)|/\log(\xi_j)$
where $\xi_j$ are uniform random numbers differing for each pole. The $j$ corresponding to the maximum of
this expression follows the desired discrete distribution. We also note that the quantity $\Re\left[r_j w(z_j)\right]$ is
exactly the Doppler broadened contribution to the integrated scattering cross section, so this sampling
procedure incurs no additional Faddeeva function evaluation overhead if this is done in tandem with a WMP cross
section lookup operation.

Finally, we emphasize that the pole sampling approximation is not
precisely consistent with the original multipole cross section representation.
Instead, it uses the fact that polynomial contributions to the cross section
negligibly affect the scattering kernel, while poles do so substantially.

\subsection{Finding the values of $\sigma_0$ and $\sigma_1$}
\label{subsection:linearization}
While it may seem that the polynomial contribution to Eq. \ref{eq:xs_form} could come as the first terms from the polynomials defined
within the windows, as \cite{biondoAlgorithmFreeGas2021} used, we have found in practice that this choice is inconsistent with the approximation of the pole sampling technique and leads to negative cross section values.

To remedy this issue, we use the heuristic that the relative error of the cross section's local approximation
is minimized by matching polynomial values in the vicinity of the dip of the resonance. The location of the
scattering resonance trough is calculated as:
\begin{equation}
  \sqrt{E}_{\text{trough}} = \frac{-b + \sqrt{b^2-abc + a^2 d}}{a}
  \label{eq:pole_dip_loc}
\end{equation}
where
\begin{align}
  a &= -\Im[\left(r_j^* - \bar{r_j^*}\right)] \\
  b &= -\Im[\left(\bar{r_j^*} p_j^* - r_j^* \bar{p_j^*}\right)] \\
  c &= -\Re[\left(\bar{p_j^*} + p_j^*\right)] \\
  d &= |p_j^*|^2 \\
\end{align}

At this point, the window index of $\sqrt{E}_{\text{trough}}$ is calculated
\footnote{The term under the square root in Eq. \ref{eq:pole_dip_loc} can sometimes be negative in the vicinity of nonphysical poles which are
artifacts of the fitting process, and dealing with imaginary quantities in this case is undesirable.
Therefore, our calculation uses a linearization of the non-pole cross section at the incident energy instead
in that case.}
. This may be a different window
from the incident neutron energy's window. The windowed multipole cross section of Eq. \ref{eq:xs_form_full}
is then evaluated at $\sqrt{E}_{\text{trough}}$ but \textit{excluding} the sampled pole $p_j$, i.e. 
\begin{equation}
  \sigma_0 = \frac{1}{E_\text{trough}} \Re \left[ \sum_{j^* \neq j \in W(E_\text{trough})} \frac{r_j}{p_j-\sqrt{E_\text{trough}}} \right] + \sum_{n=0}^{N} a_n E_\text{trough}^{n/2}
\end{equation}
From there, $\sigma_1 \approx \pd{\sigma_s(E)}{\sqrt{E}}$ is calculated at the same point, this time $\textit{only}$ including contributions
from the polynomial expansion but not from any poles. This linearization technique has been found
to improve the accuracy of our method when considering nuclides with tightly spaced resonances such as $^{235}$U.
For complete clarity, the resulting expression is:
\begin{equation}
  \sigma_1 = \sum_{n=0}^{N} a_n \frac{n-2}{2} E_\text{trough}^{n/2-2} 
\end{equation}

Finally, a linearization of the cross section in $\sqrt{E}$ space has been obtained. For use with the root finder,
the nondimensional variable $\beta (\sqrt{E} - \sqrt{E_{\text{incident}}})$ is preferable, so $\sigma_0$ is appropriately
shifted and $\sigma_1$ appropriately scaled.

\subsection{Inverting the relative speed CDF}
To sample from the relative speed PDF,
we employ the CDF inversion technique. A naive attempt at this would be a few bisection root finding
steps followed by a handful of Newton-like iterations. In practice, we've found that five bisection
iterations followed by three Halley-Newton iterations resolves the root to within acceptable tolerance;
however, a far more efficient root finder has been developed which takes a maximum of four iterations
total, only requiring more work for unusual edge cases.

The \textit{bootstrapping}
step, as we call it, is essential to an efficient implementation of MARS. The bootstrapping step
cheaply obtains an initial guess to the solution of the CDF inversion problem, from which a small number
of Newton-like iterations improve the solution.

The key to doing so lies in finding a cheap approximation to the inverse of the CDF with general
pole parameters. In order to do so, we first move from the root finding space of $x \in (-\infty, \infty)$
to the nonlinearly mapped variable $\tilde{x} = \frac{1}{2}\left(1+\erf{x}\right)$. The intuition behind using
this modified space is that as the resonances become weak and the incident neutron energy becomes high,
it can be shown that the CDF is simply equal to $\tilde{x}$ which ranges between zero and one. Resonances
and low energy free gas effects simply act as perturbations to this linear function, which enables a
good starting point for approximating the root location.

The next step in improving the CDF model in the mapped space is to observe that the contribution of
collision probability from the resonance largely does not depend on its imaginary part. Increasing
the imaginary part of the resonance broadens it and decreases its width. Therefore, the magnitude of
the jump in $w(z, x)$ when $x$ is near $\Re[z]$ quantifies the probability that the neutron experiences
a collision near the peak of the resonance. The jump in $w(z, x)$ for small $\Im[z]$ is: 
\begin{equation}
  \lim_{\Im[z]\rightarrow 0} \int_{\Re[z]-\epsilon}^{\Re[z]+\epsilon} \frac{e^{-t^2}\dif t}{z-t} = e^{-\Re[z]^2}
\end{equation}
and therefore the probability contribution due to the resonance is approximately:
\begin{equation}
  p_{\text{jump}} = \P[x\approx Re[z]] \approx \beta r_{j^*} \pi e^{-\Re[z]^2} / C
  \label{eq:approx_jump}
\end{equation}
where $C$ is the normalizing constant given by Eq. \ref{eq:normalizing_constant}.
Because this is an approximation, the probability of Eq. \ref{eq:approx_jump} may not
be bounded between zero and one, so we threshold it to that range. In practice, the
estimate provided here is accurate. We have found that this probability tends to
be added into the CDF about $\Re[z]$ over the interval 
$[\Re[z]-\frac{3}{2}\Im[z], \Re[z] + \frac{3}{2}\Im[z]]$.
This estimate could obviously be tuned for greater accuracy.

One final tool we employ to bootstrap the root finding process pertains to the values
of the CDF about $x=0$. In this case, numerous instances of functions occuring in its
expression such as $\erf(x)$ and $e^{-x^2}$ take on easily calculated values. On top of
that, the incomplete Faddeeva function has a closed form expression when $x=0$:
\begin{equation}
  w(z, 0) = \frac{1}{2} w(z) + \frac{i}{2\pi} e^{-z^2} E_1(-z^2) \quad.
\end{equation}
Because $e^{-z^2}$ is already computed and cached for the CDF inversion,
the calculation of a complex exponential integral is the only difficulty.
This is much easier and computationally cheaper to do than the more involved
$w(z, x)$ evaluation, so any off-the-shelf approximation of $E_1(-z^2)$ can
be employed here.

Addtionally, the derivatives of the CDF with respect to $x$ about $x=0$
are also easily obtainable, which we use to further improve our rootfinding
guess. So far, we have only incorporated information from the first derivative which has proven sufficient.

This leaves us with the following pieces of information from which the root
estimate is extracted: the probability due to the resonance, its width, the
value and slope of the CDF about $x=0$ i.e. $\tilde{x}=1/2$, and the
known endpoint values of the CDF at 0 and 1. We therefore construct a function
which is piecewise quadratic on the left and right of $\tilde{x}=1/2$. This
quadratic interval ranges to either the endpoints $\tilde{x}=0$ or $\tilde{x}=1$,
or the resonance's upper or lower range of probability gain, estimated here as
x$ \in [\Re[z]-\frac{3}{2}\Im[z], \Re[z] + \frac{3}{2}\Im[z]]$. Note that this
interval has to be mapped to an interval in $\tilde{x}$ space. Because the interval
of the resonance is small, the Jacobian of the transformation $x\mapsto \tilde{x}$
which is proportional to $e^{-\Re[z]^2}$ (a quantity already computed) can be
used to calculate the range in $\tilde{x}$ space.

With this knowledge, the CDF can be approximated somewhat accurately in $\tilde{x}$ space.
Despite the apparent complexity of what was just described, the inversion of the
previous paragraph's function can be done using simple branching logic and, at worst,
the solution of a quadratic equation. Because translating the inverse of the above function
into code can take nontrivial effort, C++ code to achieve this has been provisioned in
Appendix \ref{appendix:bootstrap}.
Figure \ref{fig:jump_approx} shows two examples of how this can be a quite satisfactory approximation
of the CDF in $\tilde{x}$ space when resonances are influencing the scattering distribution.
\begin{figure*}
  \centering
  \begin{subfigure}{0.49\textwidth}
    \includegraphics[width=\textwidth]{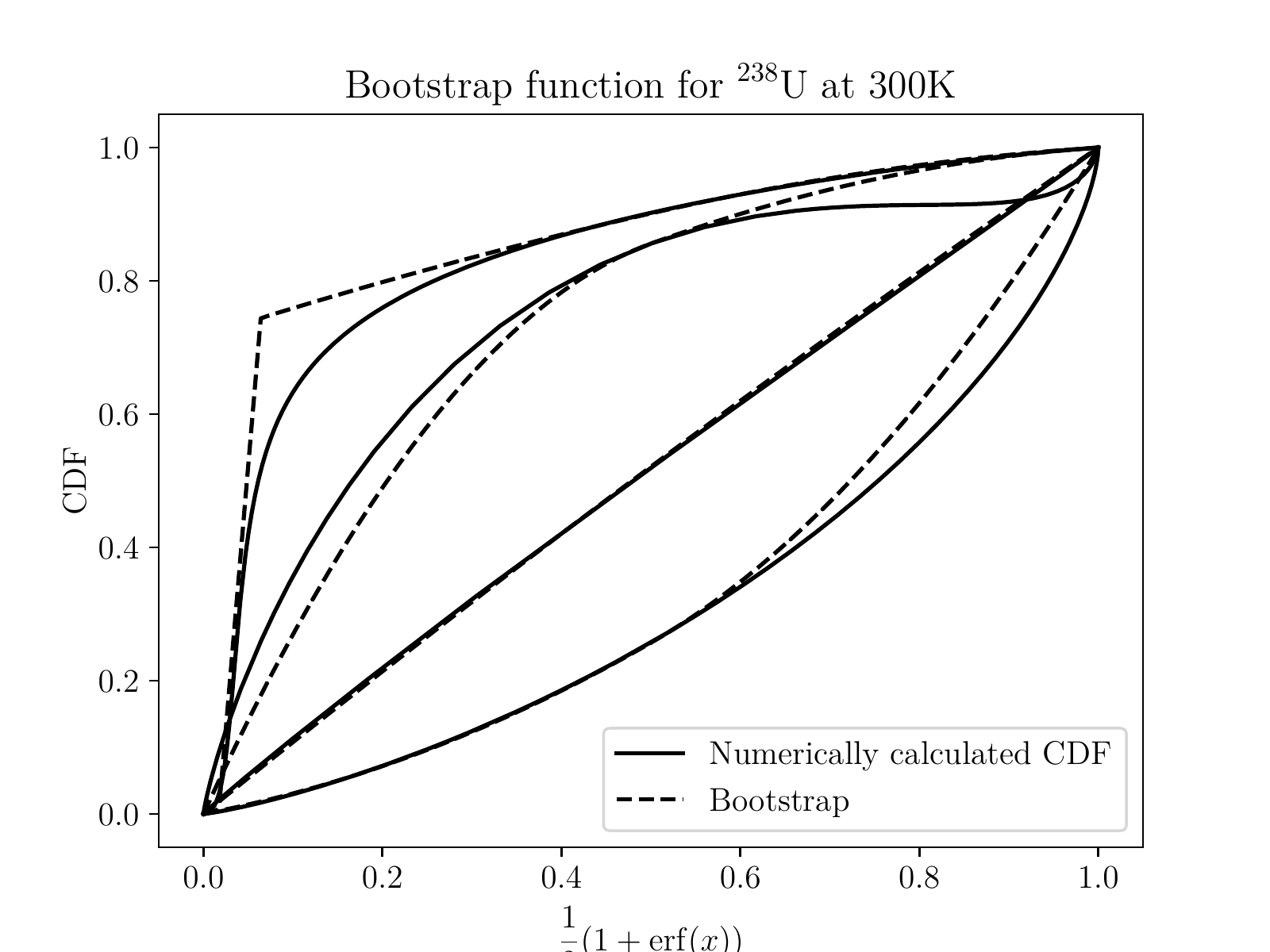}
     \caption{$T = 300$ kelvin}
  \end{subfigure}
  \begin{subfigure}{0.49\textwidth}
    \includegraphics[width=\textwidth]{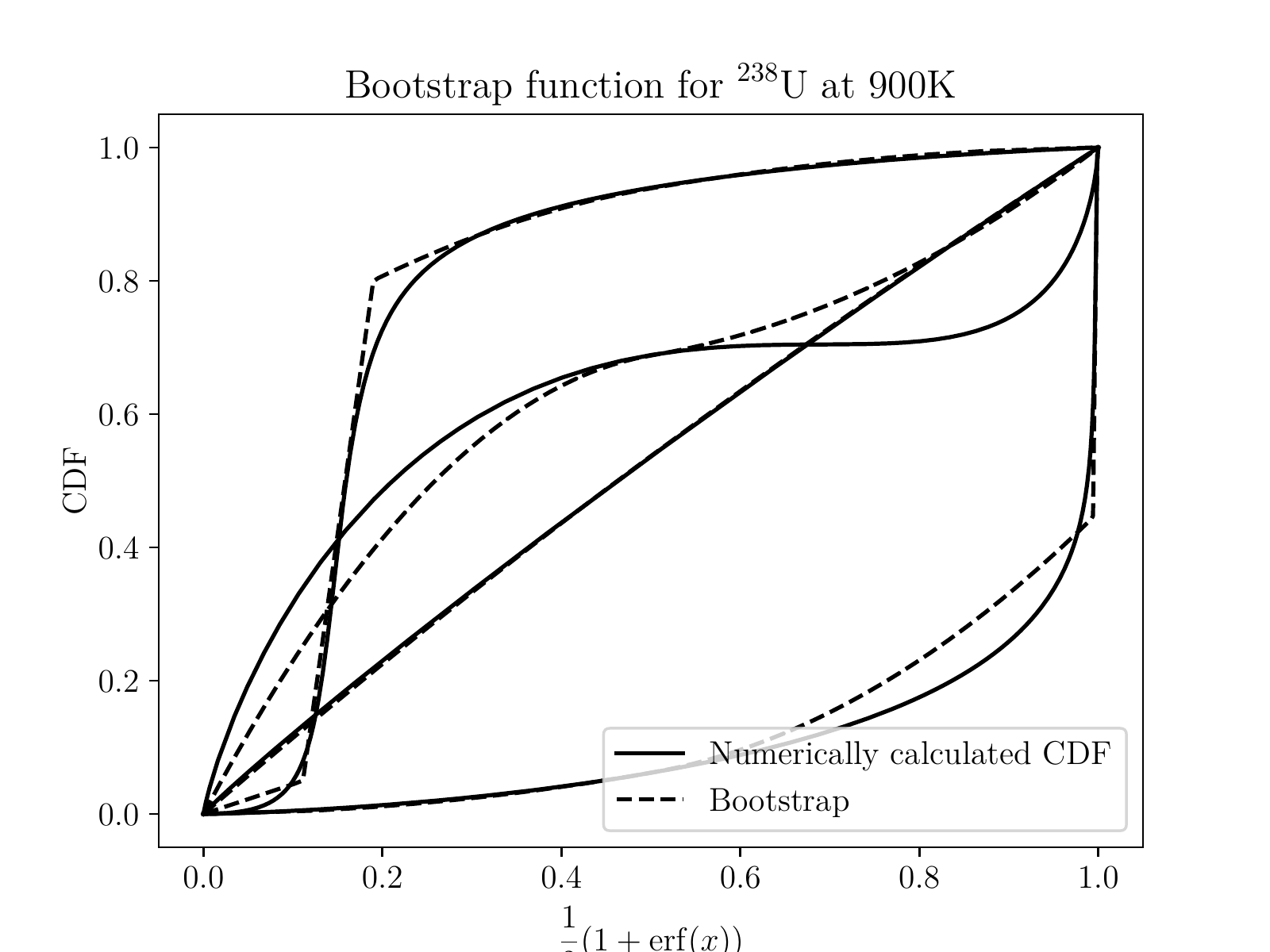}
     \caption{$T = 900$ kelvin}
  \end{subfigure}
  \caption{The bootstrapping CDF provides a fairly accurate, easily invertible approximation
  to the true relative speed CDF to kickstart the root finding process. The pairs of lines,
  moving from top to bottom, represent 35.25, 36.25, 38.25, and 66.25 eV incident neutron energies.}
  \label{fig:jump_approx}
\end{figure*}

\section{Results}

\subsection{Calculation of $w(z, x)$}
\begin{figure}
   \centering
   \begin{subfigure}[b]{0.45\textwidth}
       \centering
       \includegraphics[width=\textwidth]{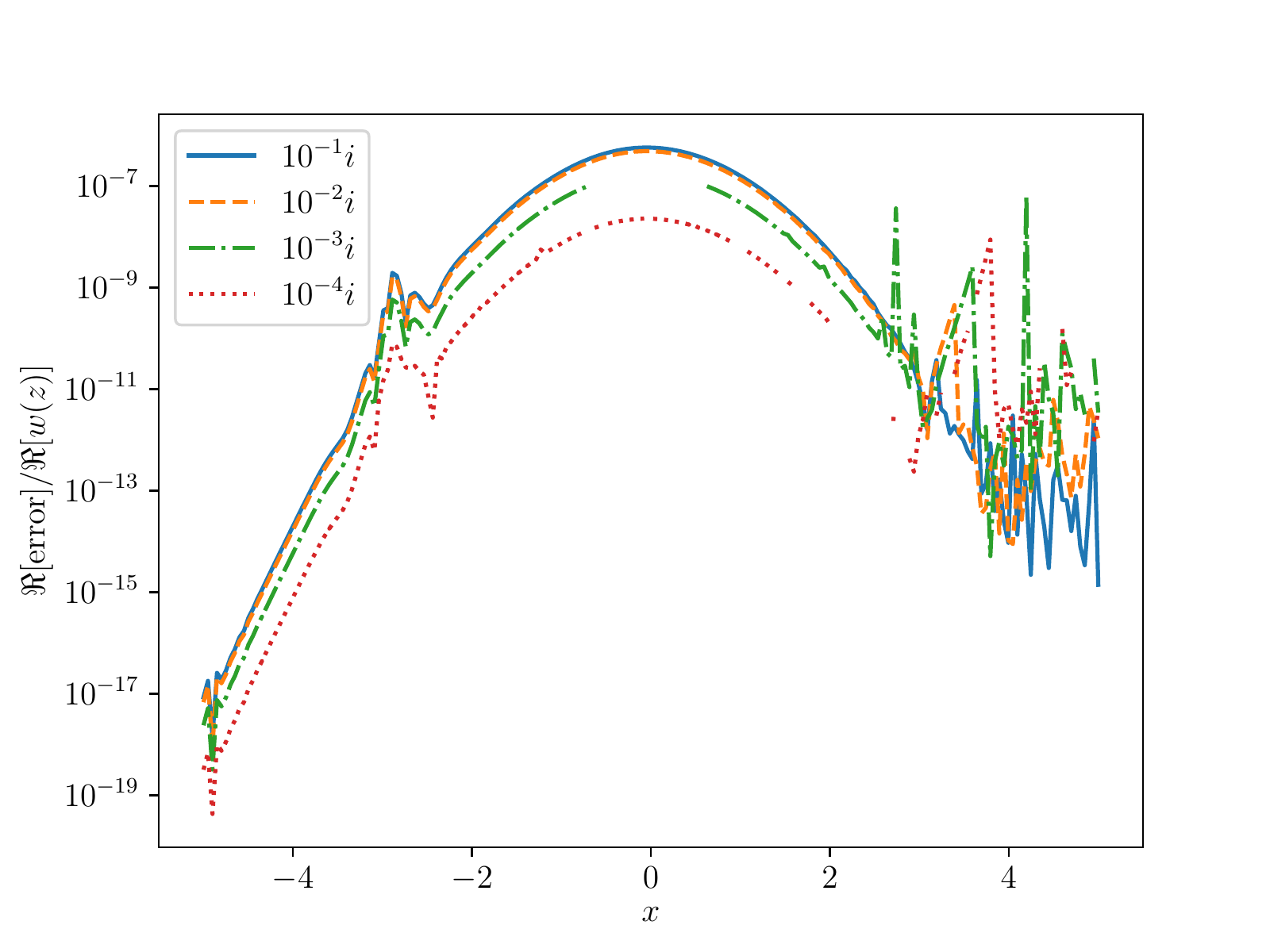}
       \caption{$\Re[z]=-2.9$}
   \end{subfigure}
   \begin{subfigure}[b]{0.45\textwidth}
       \centering
       \includegraphics[width=\textwidth]{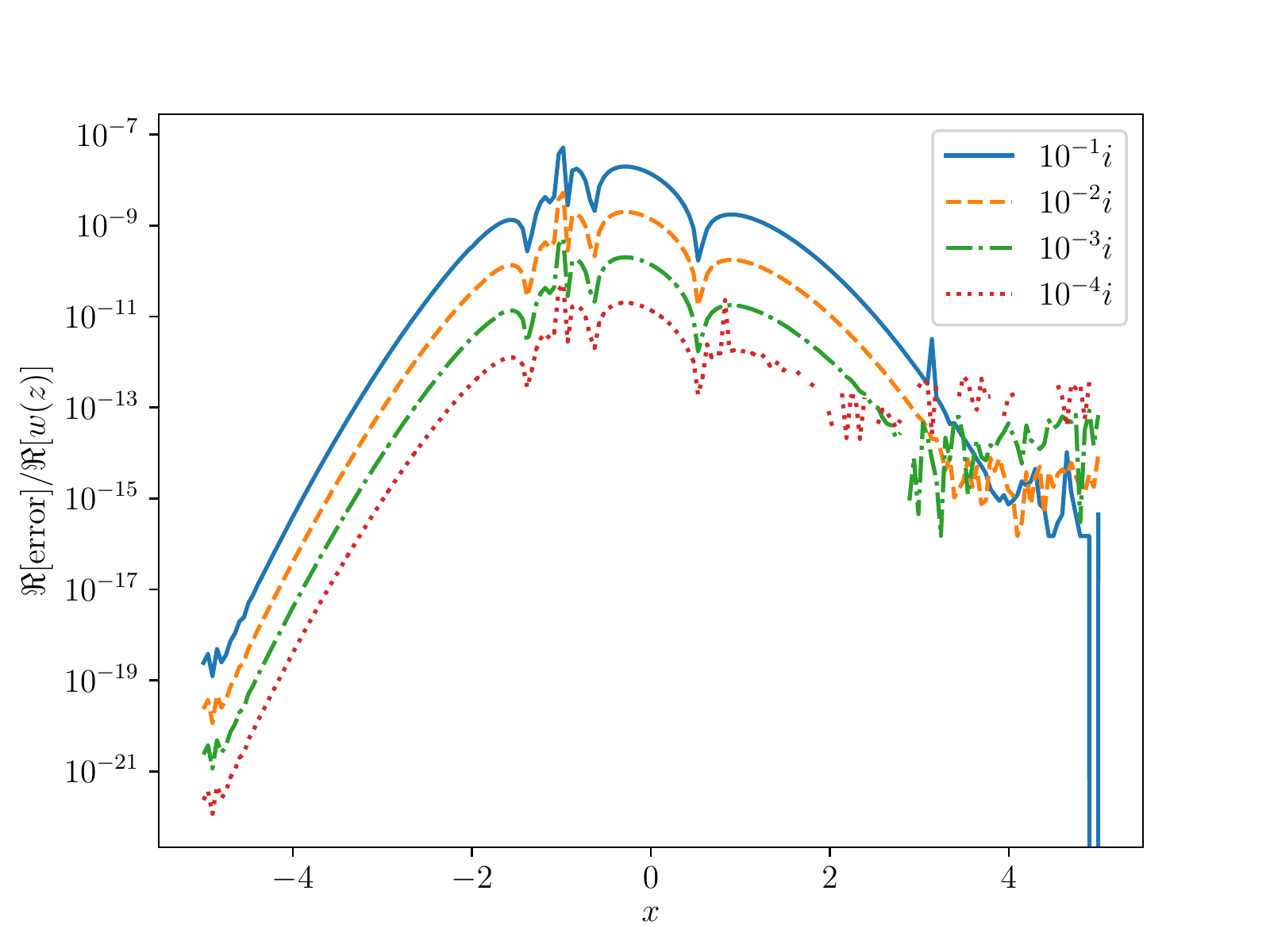}
       \caption{$\Re[z]=-1.0$}
   \end{subfigure}

   \begin{subfigure}[b]{0.45\textwidth}
       \centering
       \includegraphics[width=\textwidth]{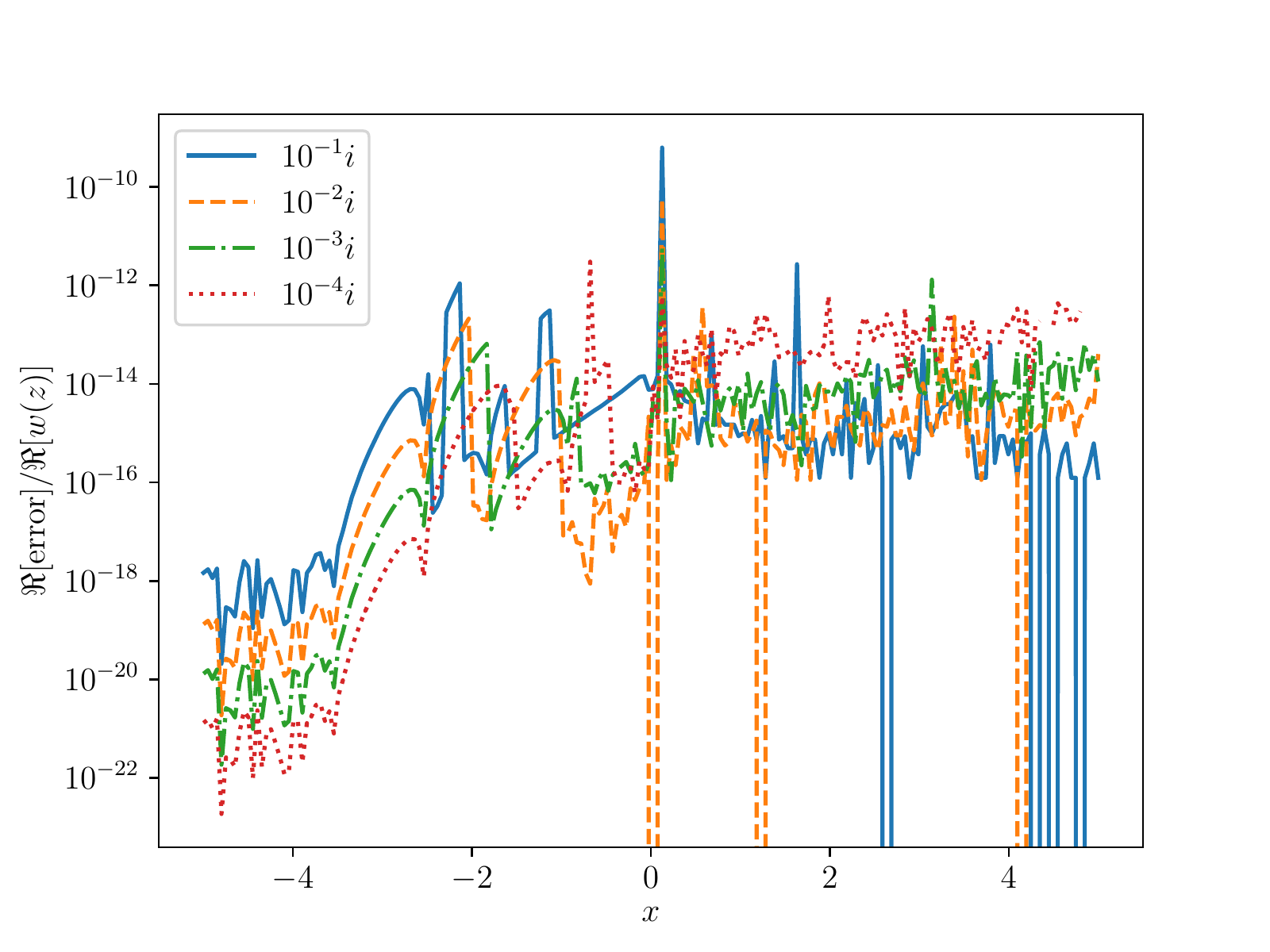}
       \caption{$\Re[z]=0.0$}
   \end{subfigure}
   \begin{subfigure}[b]{0.45\textwidth}
       \centering
       \includegraphics[width=\textwidth]{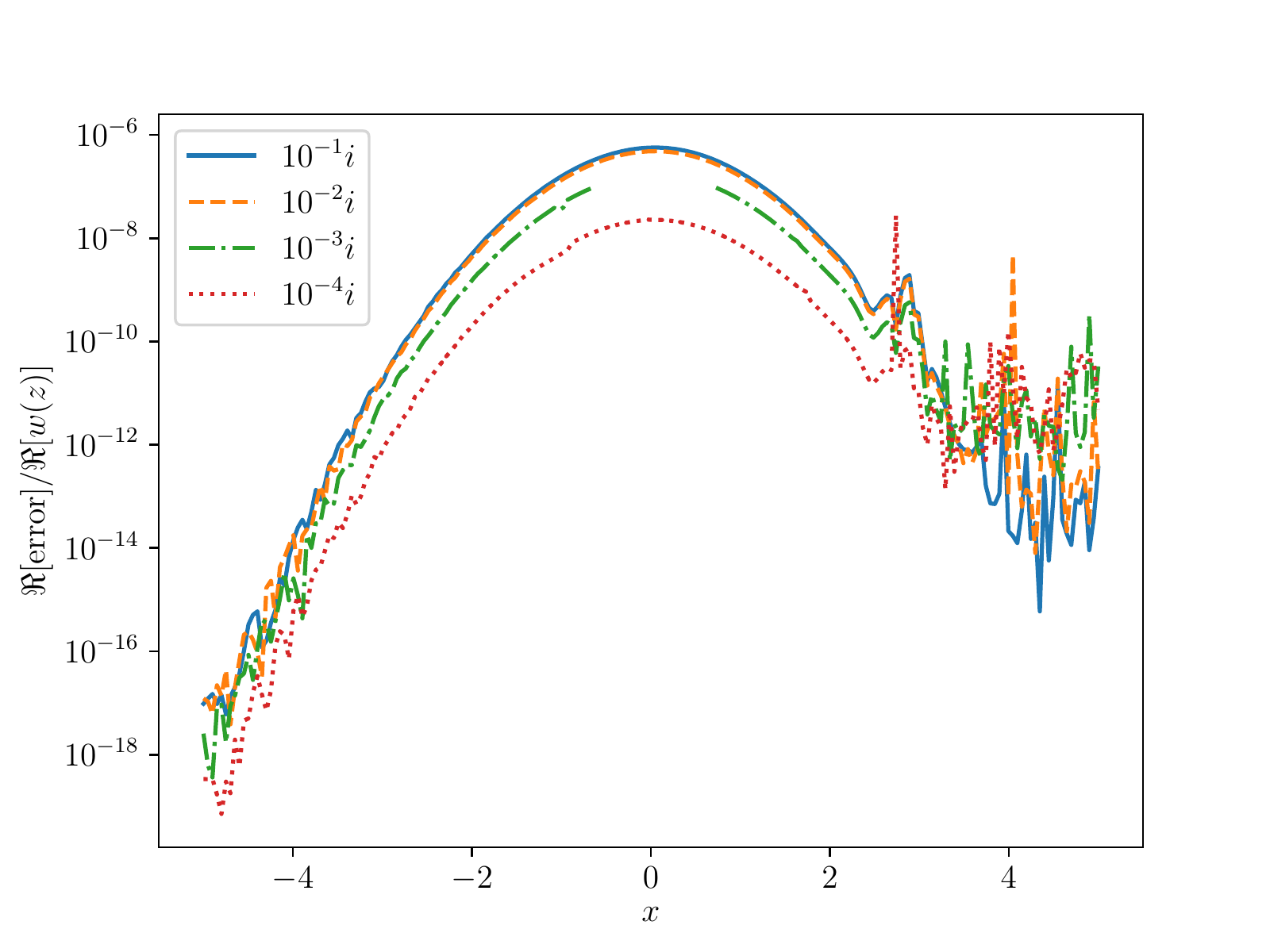}
       \caption{$\Re[z]=2.9$}
   \end{subfigure}

   \caption{Error of $\Re[w(z, x)]$ for a few values of $z$. The legend is the imaginary number added to the real part specified in each figure's caption.
   The plotted error, $w_{\text{approx}}(z, x)-w(z, x)$ is normalized by $\Re[w(z)]$ to match the scaling of Fig. \ref{fig:inc_fad_rp_line_plots}.}
   \label{fig:inc_fad_err_line_plots}
\end{figure}

In order to test the accuracy of Alg. \ref{alg:inc_fad_alg}, we haved computed reference values of $w(z, x)$
using \texttt{scipy}'s \cite{virtanenSciPyFundamentalAlgorithms2020} adaptive quadrature routine, \texttt{scipy.integrate.quad}, to evaluate the integral
formulation Eq. \ref{eq:inc_fad}. In approximation of the jump integral Eq. \ref{eq:jump_integral}, only the first five terms in the series are retained.
Where functions such as $\log(x)$ or $e^x$ appear, C++ standard library implementations have been employed.
The implementation of $w(z)$ from \cite{johnsonFaddeevaPackage} has been employed.
This results in the error profiles exhibited by Fig. \ref{fig:inc_fad_err_line_plots}, where we have plotted
the real part of $\left(w_{\text{approx}}(z, x) - w(z, x)\right) / w(z)$. Because only the real part is of interest
in resonance upscatter calculations, results on the imaginary component's error are omitted.

\subsection{Single Energy Testing}
\begin{figure*}
  \centering
  \begin{subfigure}{0.49\textwidth}
    \includegraphics[width=\textwidth]{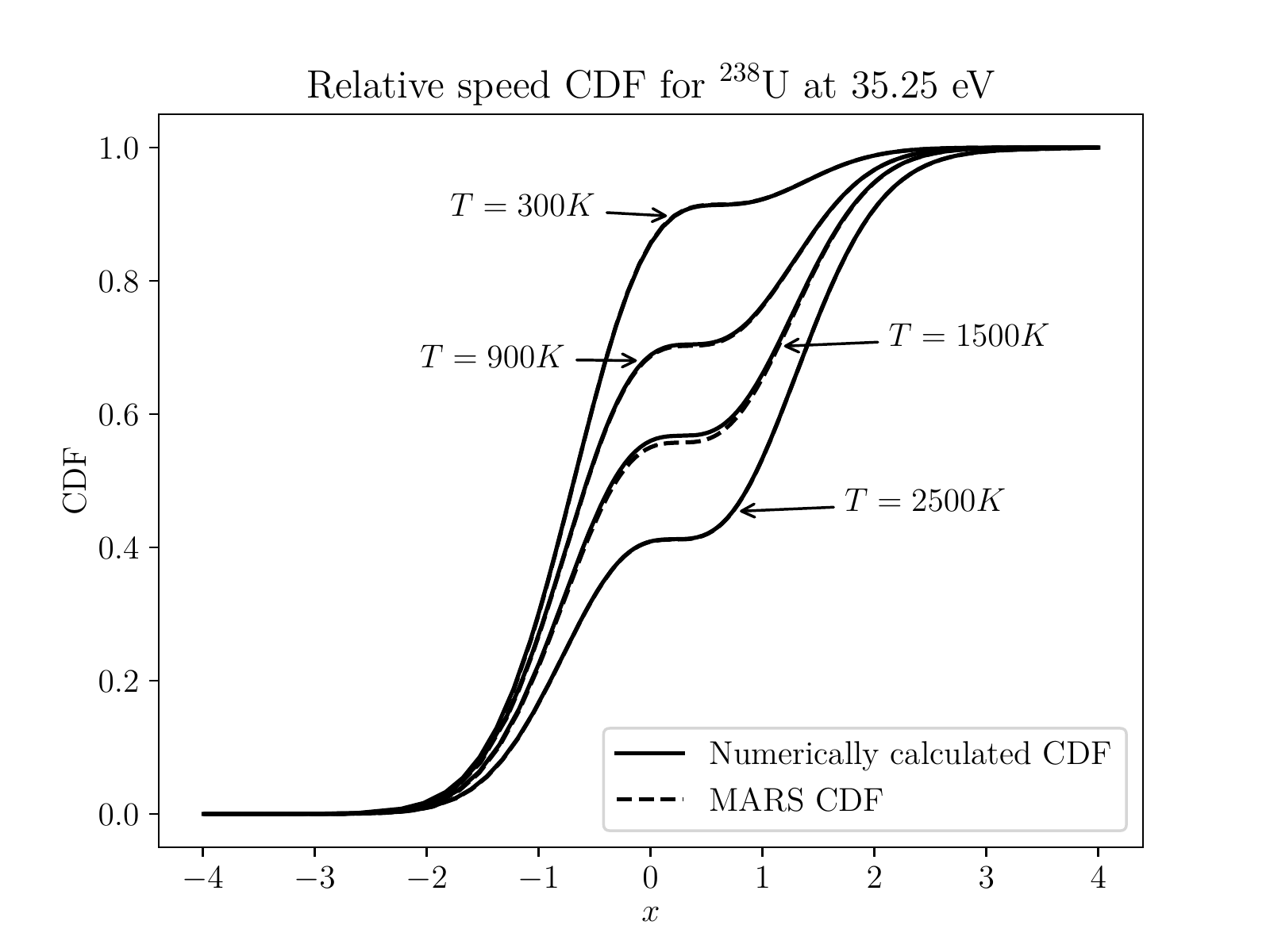}
     \caption{35.25 eV}
  \end{subfigure}
  \begin{subfigure}{0.49\textwidth}
    \includegraphics[width=\textwidth]{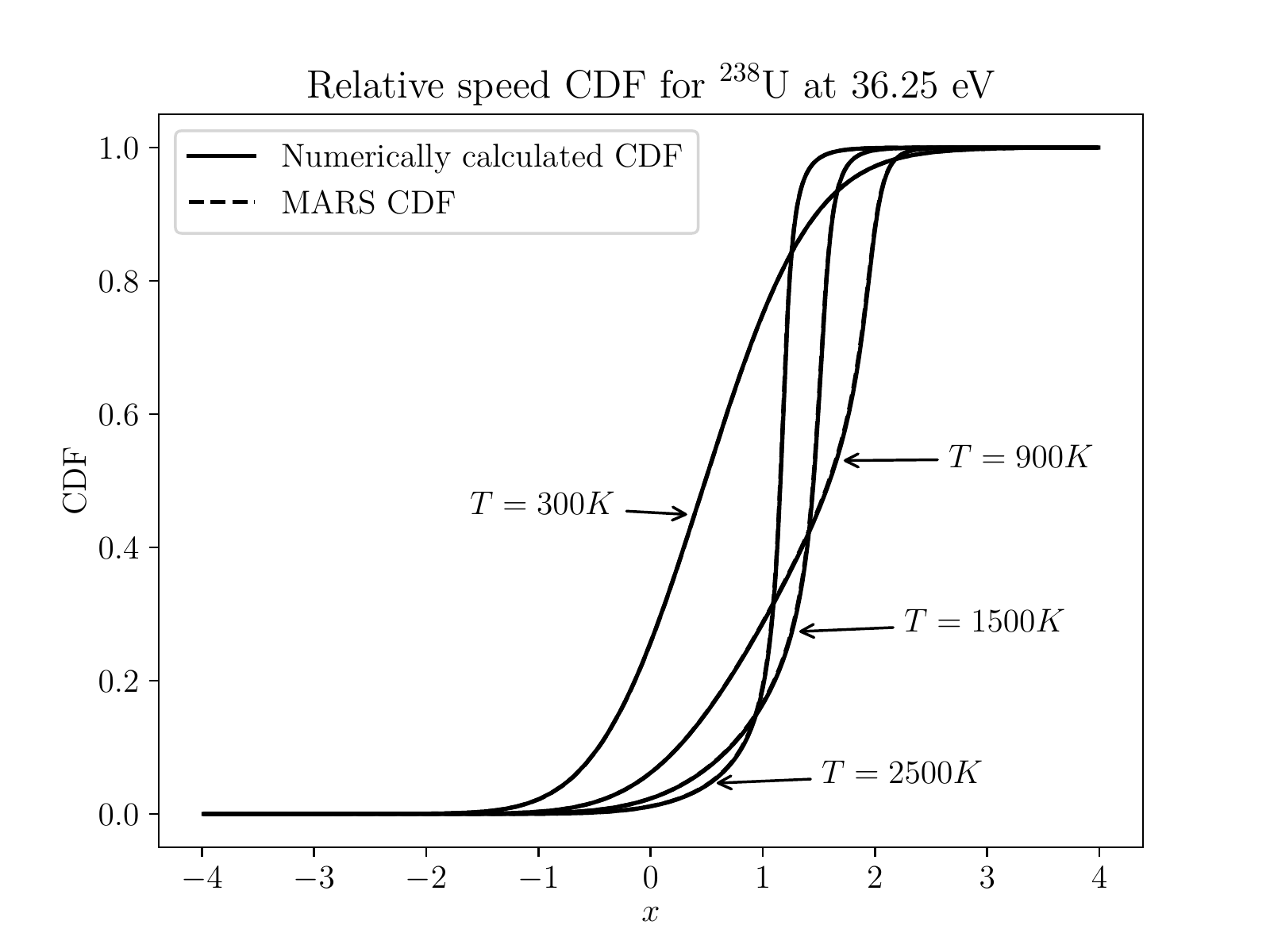}
     \caption{36.25 eV}
  \end{subfigure}
  \caption{The MARS analytic CDF matches numerically integrated relative speed
    cumulative distributions. Some error can be observed for the 1500K case at 35.25 eV;
    scattering in the resonance dip is fortunately an extremely rare event.}
  \label{fig:cdf_compare}
\end{figure*}

We first present in Fig. \ref{fig:cdf_compare} the relative
speed distribution of $^{238}$U for two different energies and a few temperatures as calculated both
by numerical integration and the MARS analytic CDF. The energies correspond to being in the trough
and near the peak of a scattering resonance.
These plots clearly
show the influence of the resonances on the double differential cross section; a nuclide with constant
cross section has a relative speed distribution which is very nearly an error function at epithermal
energies. The relative
speed distribution near resonances has a jumping effect which is governed by $w(z, x)$.
They bear a resemblance to $w(z, x)$ behavior depicted by Fig. \ref{fig:inc_fad_rp_line_plots}.

\begin{figure*}
  \centering
  \begin{subfigure}{0.49\textwidth}
    \includegraphics[width=\textwidth]{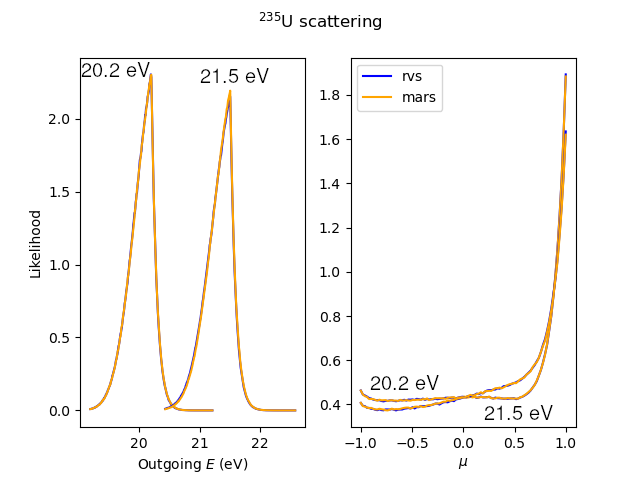}
     \caption{$^{235}$U}
  \end{subfigure}
  \begin{subfigure}{0.49\textwidth}
    \includegraphics[width=\textwidth]{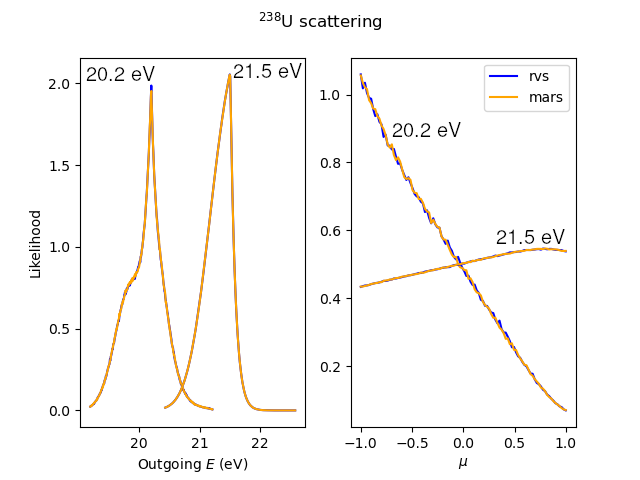}
     \caption{$^{238}$U}
  \end{subfigure}
  \caption{Scattering at 1200K matches results from RVS method well at two different energies. These energies interact with resonances for both nuclides.}
  \label{fig:rvs_compare}
\end{figure*}

If the relative speed distribution is correct, the resultant double-differential scattering
distribution is also correct. Fig. \ref{fig:rvs_compare} shows this is the case for our method
when compared to the RVS method of \cite{romanoImprovedTargetVelocity2018}. These results were
obtained from our modified version of OpenMC, available at \texttt{github.com/gridley/openmc/tree/mars}.
It also shows that the pole sampling technique successfully works for $^{235}$U and its tightly spaced
resonances.

\subsection{Pin Cell Reactivity Feedback}
\begin{figure}
  \centering
  \includegraphics[width=0.7\textwidth]{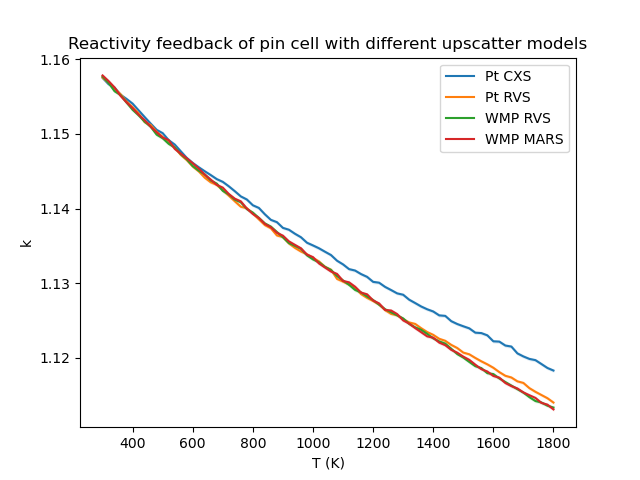}
  \caption{MARS matches the $k$ eigenvalue of the RVS method on a 2.4\% enriched fresh PWR pin cell problem.
  Line width represents estimated standard deviation of the mean.}
  \label{fig:reactivityfeedback}
\end{figure}

The 2.4\% enriched PWR pin cell example from OpenMC's suite of example problems was used to calculate
Doppler reactivity feedback effects with four different models. The first and second
used pointwise cross sections that were interpolated between 300, 600, 900, 1200, and 2500 kelvin.
The model was run at temperatures ranging from 300 to 1800 kelvin in increments of 20 kelvin.
Of the two using pointwise cross sections, one used the historical constant cross section free gas
scattering approximation, and the second used the RVS method. The second two cases both used windowed
multipole cross sections, one using RVS and the second MARS. The ENDFB-VII.1 nuclear dataset was employed.
Figure \ref{fig:reactivityfeedback} shows how these cases compare. Two hundred cycles with ten inactive
were employed, using 200,000 particles per cycle. $k_{\text{eff}}$ was thus converged to 20 pcm for each case.

It can be seen that the pointwise cross section representation incurs some interpolation error
between 1200 and 1800 kelvin. The MARS method matches the RVS results where multipole cross sections
were employed. We can thus conclude that the new method works correctly across the range of energies
where resonances influence the double-differential cross section at temperature for nuclides of both
strong, distantly spaced resonances ($^{238}$U) and closely spaced weak resonances ($^{235}$U).

\subsection{Influence on Tracking Rate}
Finally, in order to determine the computational efficiency of the new method,
tracking rate comparisons were carried out on the same PWR pin cell example problem. 
The computational performance of both inactive and active cycles was assessed. For the
active cycles, a 100x100 Cartesian mesh tallied flux, fission rates, and neutron production rates using track-length estimators. In addition, a spatially homogenized energy spectrum
tally consisting of 500 equal lethargy bins was applied.

An Intel Xeon W-2133 with six physical cores carried
out the calculations, and obtained the results depicted in Table
\ref{tab:speed_result}. This clearly demonstrates the computational efficiency of MARS compared to the RVS and DBRC methods. While it does not outperform RVS in this scenario,
future work will explore its performance on vector computer architectures where we expect
it to outperform.

\begin{table}
    \centering
    \begin{tabular}{c|c|c}
         Method & Inactive & Active \\
         \hline
         CXS  & 60.5 & 11.4 \\
         DBRC & 57.0 & 11.1 \\
         RVS  & 58.3 & 11.3 \\
         MARS & 60.3 & 11.1 \\
    \end{tabular}
    \caption{Tracking rate in thousand particles per second obtained by the constant cross section treatment, and three resonance upscatter models.
    MARS is comparable in speed to widely accepted techniques.}
    \label{tab:speed_result}
\end{table}

The computational expense incurred by tallying tends to render the performance impact
of our new method particularly negligible. Collision estimators could be used on
the mesh tally to improve the tracking rate, but we arbitrarily opted for track length estimators. Due to subtle hardware-related effects such
as cache utilization or branch prediction, 
the tracking rates of the three resonance upscatter handling methods have different
relative performances when comparing active and inactive cycles. Future work will explore detailed performance results on a variety of architectures.

\section{Conclusion}

The multipole formalism carries a variety of advantages compared to pointwise cross sections.
Aside from its potential gains in computational efficiency on modern compute architectures,
it enables accurate Doppler broadening without a library size tradeoff \cite{joseyWindowedMultipoleCross2016}, elegant sensitivity
quantification, and narrows the gap between R matrix theory and the cross 
section representation \cite{ducruWindowedMultipoleRepresentation2021}. This work develops yet another advantage to the windowed multipole
formalism: closed-form resonance upscatter treatment.

We have demonstrated that the new method matches the results obtained by other
resonance upscatter techniques. To achieve this, we derived an expression
for the target relative speed distribution, and identified a novel special function
which universally arises in this application. Novel numerical techniques that balance
efficiency and accuracy were derived, implemented, and tested. The overall scheme
was shown to achieve the same tracking rate as other resonance upscatter modeling methods.

The new method called multipole analytic resonance scattering (MARS) overcomes the storage
requirements of relative speed tabulation \cite{choiOptimizationNeutronTracking2021}, and avoids rejection sampling as employed by other common approaches. Without
a need to access intermediate storage, the accesses to global memory can be reduced on GPU architectures.
On top of that, the work discrepancy between threads incurred by rejection sampling  on GPUs is similarly
overcome. Future work will explore the implementation and optimization of this method on GPUs.

%

\section{Acknowledgements}
This work was partially supported by the U.S. Department of Energy through the Los Alamos National Laboratory. Los Alamos National Laboratory is operated by Triad National Security, LLC, for the National Nuclear Security Administration of U.S. Department of Energy (Contract No. 89233218CNA000001).
This material is also based upon work partially supported under an Integrated University Program Graduate
Fellowship. 
This research was also partially supported by the Exascale Computing Project (17-SC-20-SC), a collaborative effort of the U.S. Department of Energy Office of Science and the National Nuclear Security Administration.

Any opinions, findings, conclusions or recommendations expressed in this publication are
those of the author(s) and do not necessarily reflect the views of the Department of Energy
Office of Nuclear Energy.

\printbibliography

\appendix
\section{Derivation of Eq. \ref{eq:inc_fad_identity}}
The forthcoming discussion has not been made mathematically rigorous for sake of brevity and
the context of a nuclear engineering journal.
We begin by defining the auxiliary complex function F(z):
\begin{equation}
  F(z) = e^{z^2} w(z, x)
\end{equation}

The complex line integral theorem can then be applied when $\Im[z] > 0$:
\begin{equation}
  F(z) - F(0) = \int_0^z \od{F}{z}|_{z=z'} \dif z'
\end{equation}
Computing $\od{F}{z}$ and inserting then reveals:
\begin{equation}
  e^{z^2} w(z, x) - w(0, x) = \int_0^z \left(2 z' e^{z'^2} w(z', x) - e^{z'^2}\frac{i}{\pi}  \int_{-\infty}^x \frac{e^{-t^2}\dif t}{(z'-t)^2 }\right) \dif z'
\end{equation}
where the linearity of integration has been employed, and the interchange of differentiation and integration has also been used.
The innermost integrals can now be computed exactly, carrying the $z'$ through to the integral defining the incomplete Faddeeva function.
This results in:
\begin{equation}
  e^{z^2} w(z, x) - w(0, x) = \frac{i}{\pi} \left(\int_0^z  \frac{e^{z'^2} e^{-x^2}}{x-z'}\dif z' - \sqrt{\pi} (1+\erf(x)) \int_0^z e^{z'^2} \dif z' \right)
  \label{eq:intermediate_inc_fad_2}
\end{equation}

Recalling that the Faddeeva function can be defined as
\begin{equation}
  w(z) = e^{-z^2} \left(1+\frac{2i}{\sqrt{\pi}} \int_0^z e^{t^2} \dif t\right) \quad,
\end{equation}
we can identify $w(z)$ as the trailing term of Eq. \ref{eq:intermediate_inc_fad_2}. The
term $w(0, x)$ must be interpreted in a principal value sense, which results in a contribution
in the form of a Heaviside function. The following expression then results:
\begin{multline}
  w(z, x) = e^{-z^2}\left( -\frac{1}{2} (\Ei(-x^2) + \frac{i}{\pi} e^{-x^2} \int_0^z \frac{e^{z'^2}}{x-z'}\dif z' \right) + \\
  \frac{1}{2}(1+\erf(x))(w(z) - e^{-z^2}) + h(x) e^{-z^2} \quad.
  \label{eq:intermediate_inc_fad}
\end{multline}
This result could perhaps be used for numerical calculations of $w(z, x)$. However, it suffers the shortcoming that
the exponential integral term goes to infinity for $x=0$, which is cancelled out by the integral term. However, $w(z, x)$ 
is well-defined at $x=0$, and the addition of branching logic to numerical routines to handle this case would be cumbersome.
The expression can be made more amenable to numerical approximation with some further simplification.

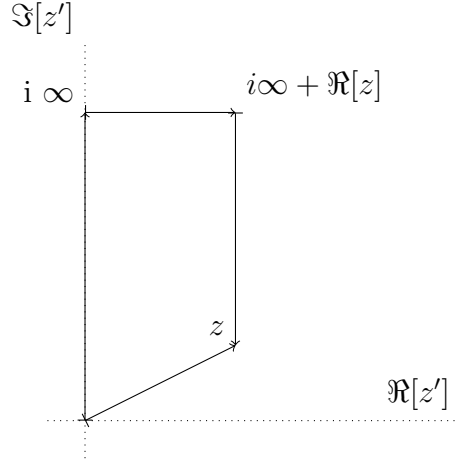
\begin{figure}
  \centering
\begin{tikzpicture}
  \begin{scope}{thick,font=\scriptsize}
    \draw [dotted] (-0.5, 0) -- (5, 0) node [above left] {$\Re[z']$};
    \draw [dotted] (0, -0.5) -- (0, 5) node [above left] {$\Im[z']$};
    \draw [|->] (0, 0) -- (2, 1) node [above left] {$z$};
    \draw [|->] (0, 0) -- (0, 4.1) node [above left] {i $\infty$};
    \draw [|->] (0, 4.1) -- (2, 4.1) node [above right] {$i \infty+\Re[z]$};
    \draw [|->] (2, 4.1) -- (2, 1) node [above left] {};
  \end{scope}
\end{tikzpicture}
\caption{Modified contour used to cancel exponential integral in Eq. \ref{eq:intermediate_inc_fad}. The contribution from the top line is zero.}
\label{fig:modded_path}
\end{figure}

The integral term goes from 0 to $z$, and the integrand encloses no poles of the following path
whenever $x\neq 0$. As such, a change of integration path is employed: the contour of Fig. \ref{fig:modded_path}
results in a convenient cancellation of terms. The top leg of the contour is zero from the $e^{t^2}$ term, resulting in:
\begin{equation}
  \int_0^z \frac{e^{t^2}}{x-t} \dif t = \int_0^{i \infty} \frac{e^{t^2}}{x-t} \dif t - \int_z^{i \infty} \frac{e^{t^2}}{x-t} \dif t
  \label{eq:contour_equiv}
\end{equation}

A little bit of algebra shows that:
\begin{equation}
\int_0^{i \infty} \frac{e^{t^2}}{x-t} \dif t = \frac{1}{2} e^{x^2} \left(-i \pi (\erf(x)-\text{sign}(x)) + \Ei(-x^2) \right) 
\end{equation}
The sign function term ends up cancelling out the Heaviside term upon substitution of Eq. \ref{eq:contour_equiv} back to Eq. \ref{eq:intermediate_inc_fad}.
The second term in Eq. \ref{eq:contour_equiv} easily can be transformed via the change of variables $t'=-it$ to the integral in Eq. \ref{eq:inc_fad_identity},
thus yielding Eq. \ref{eq:inc_fad_identity}.

\section{Derivation of Eq. \ref{eq:asymp_nstar}}
\label{appendix:asymp_nstar}

The goal is to find the smallest $n$ such that
\begin{equation}
  \frac{x^{n} E_1(x)}{n! E_{n+1}(x)} < 1
  \label{eq:solvethis}
\end{equation}
The variation of the left hand side function of Eq. \ref{eq:solvethis} as $n$ increases, for a constant value of $x$, has been shown to be first increasing from one,
reaching a maximum, and monotonically decreasing from that point, never becoming negative \cite{gautschiRecursiveComputationCertain1961}.
Firstly, the exponential integrals are replaced with the equivalent upper incomplete gamma function:
\begin{equation}
  E_n(x) = x^{n-1} \Gamma(1-n, x)
\end{equation}
so the equation becomes:
\begin{equation}
  \Gamma(0, x) = n! \Gamma(-n, x)
\end{equation}
Using the asymptotic formula for the upper incomplete gamma function $\Gamma(s, x) \rightarrow x^{s-1} e^{-x}$ gives
this approximation to solve:
\begin{equation}
  x^n \approx n!
\end{equation}
Which can be solved approximately by first inserting Stirling's formula:
\begin{equation}
  x^n \approx \sqrt{2 \pi n} \left(\frac{n}{e} \right)^n 
\end{equation}
Taking the $n$th root results in 
\begin{equation}
  e x \approx (2 \pi n)^{\frac{1}{2n}} n
\end{equation}
the second term on the right can be approximated by expanding the exponential, for large $n$:
\begin{equation}
  e x \approx (1+\frac{1}{2n} \log(2 \pi n)) n
\end{equation}
This can be solved exactly in terms of the Lambert W function:
\begin{equation}
  n \approx \frac{1}{2}W\left(\frac{e^{2 e x}}{\pi}\right)
\end{equation}
The asymptotic property of the Lambert
W function that $W(z) \approx \log z - \log \log z$ is then used to obtain Eq. \ref{eq:asymp_nstar}.

\section{Asymptotic Approximation for $R(m, z)$ of Eq. \ref{eq:r_integral}}
\label{appendix:r_integral_asymptotic}
An efficient approximation can be found by grouping the integrand as:
\begin{equation}
  R(m, z) = \int_0^\infty f^{(0)}(t, m) \sin(2z t)\dif t = -\frac{1}{2z} \int_0^\infty f^{(0)}(t, m) \pd{}{t} \left[\cos(2zt) \right] \dif t
  \label{eq:rint_ibp}
\end{equation}
where
\[
  f^{(0)}(t, m) = \frac{e^{-t^2}}{t^2 + m^2} \quad.
\]
Integrating the rightmost expression expression in Eq. \ref{eq:rint_ibp} by parts repeatedly
results in a divergent series approximation of the form:
\begin{equation}
  R(m, z) = \frac{f^{(0)}(0, m)}{2 z} + \frac{f^{(2)}(0, m)}{8 z^3} + \frac{f^{(4)}(0, m)}{32 z^5} + \frac{f^{(6)}(0, m)}{128 z^7} + \cdots
  \label{eq:rint_asymp_series}
\end{equation}
where $f^{(n)}(t, m)$ denotes the $n$th derivative of $f^{(0)}(m, t)$ with respect to $t$.
Some of the subsequent values evaluated about $t=0$ are:
\begin{equation}
  f^{(2)}(0, m) = -\frac{2(1+m^2)}{m^4}
\end{equation}
\begin{equation}
  f^{(4)}(0, m) = \frac{12(2+2m^2+m^4)}{m^6}
\end{equation}
\begin{equation}
  f^{(6)}(0, m) = -\frac{120(6+6m^2+3m^4+m^6)}{m^8}
\end{equation}
\begin{equation}
  f^{(8)}(0, m) = \frac{1680(24+24m^2+12m^4+4m^6+m^8)}{m^{10}}
\end{equation}
While seemingly progressing without a clear pattern, after a considerable
amount of staring at these expressions, a simple recursive formula can be 
obtained to compute these values: prime for computer implementation.
Consider the sequences $a_n, c_n \in \mathbb{R}$ defined by
\begin{equation}
  a_0 = m^{-2}; \quad c_0 = 2
\end{equation}
and
\begin{equation}
  a_{n+1} = \frac{2 n (2n-1) a_n + c_n}{m^2}; \quad c_{n+1} = (4n+2)c_n \quad.
\end{equation}
Using this, one can show that $f^{(2 n)}(0, m) = a_n$. This allows for easy
evaluation of the asymptotic series of Eq. \ref{eq:rint_asymp_series}.
Numerical experimentation has shown this to be an excellent approximation with
a maximum error around $10^{-6}$ when $|z|=5$ and $|m|=1$, retaining only five terms.
The error rapidly falls from there as $|z|\rightarrow \infty$ and $|m|\rightarrow \infty$.

\section{Asymptotic Approximation for $J(z, x)$ of Eq. \ref{eq:jump_integral}}
\label{appendix:jump_asymptotic}
Compared to the asymptotic approximation for the $R(m, z)$ integral, a clean
expression for simple computer code is not available to our knowledge. Obtaining an
asymptotic expression thus relies on access to a computer algebra system. Finding
this starts by applying a simple change of variables to Eq. \ref{eq:jump_integral}
to find:
\begin{equation}
  J(z, x) = \int_0^{\Im[z]} \frac{e^{-t^2} e^{2i\Re[z]t} \dif t}{i(\Re[z]-x)-t}
\end{equation}
Where it becomes clear that numerical difficulty from expanding the exponential term
originates from the $e^{2i\Re[z]t}$ modulation. This is the term to isolate to obtain
the correct asymptotic behavior as the integrand becomes increasingly oscillatory.
The standard repeated integration by parts procedure can then be applied. This is pure
tedium, so we simply report the C++ code which evaluates five terms below.
\begin{lstlisting}
const std::complex<double> pp1 = 6.0 + m2*(6.0 + 3.0*m2) + 
  zr*(m*(-3.0 - 3.0*m2) +
  zr*(m2*(2.0 + 2.0*m2) +
  zr*(-2.0*m2*m + 4.0*m2*m2*zr))) +
  zi*(zr*(3.0*ii + m2* (3.0*ii - 6.0*ii*m2) + 
  zr*(m*(-4.0*ii - 4.0*ii*m2) +
  zr*(m2*(6.0*ii - 4.0*ii*m2) - 16.0*ii*m2*m*zr))) +
  zi*(6.0 + m2*(6.0
  - 12*m2) + zr*(m*(-3.0 - 18.0*m2) + 
  zr*(-2 - 4.0*m2*m2 + zr*(m*(6.0
  - 16.0*m2) - 24.0*m2*zr))) + zi*(40.0*ii*m2*m + zr*(3.0*ii
  + m2*(18.0*ii + 4.0*ii*m2) + zr*(m*(-4.0*ii + 16.0*ii*m2) +
  zr*(-2.0*ii + 24.0*ii*m2 + 16.0*ii*m*zr))) +
  zi*(3.0 + m2*(48.0 +
  4.0*m2) + zi*(m*(-24.0*ii - 16.0*ii*m2) +
  zi*(-4.0 - 24.0*m2 +
  zi*(16.0*ii*m + 4.0*zi + 4.0*ii*zr) +
  (-16.0*m - 4.0*zr)*zr) +
  zr*(-24.0*ii*m2 + (-16.0*ii*m - 4.0*ii*zr)*zr)) +
  zr*(m*(6.0 +
  16.0*m2) +
  zr*(-2.0 + 24.0*m2 + zr*(16.0*m + 4.0*zr)))))));
const double pp2 = (-6.0 + m2*(-6.0 - 3.0*m2) +
  zr*(m*(3.0 + 3.0*m2) + zr*(m2*(-2.0 -
  2.0*m2) + zr*(2.0*m2*m - 4.0*m2*m2*zr))))/(m2*m2*m);

// Note:std::exp(zi*(zi - 2.0*ii*zr)) = exp(-z^2) / exp(-zr^2)
result =
  (pp2 + pp1 / (cache.emz2 / cache.emrz2 * std::pow(m - ii * zi, 5))) /
  (8. * std::pow(zr, 5));
\end{lstlisting}
\section{Root Finding Bootstrap Function}
\label{appendix:bootstrap}
\begin{lstlisting}
double rootfinding_bootstrap_guess(double xi,
         double apprx_0_cdf, // approx CDF at x=0
         double dcdx, // approx PDF at x=0
         double jump, // probability jump at resonance
         std::complex<double> z) {

  // Note: can approximate length as e^{-z^2} * 3 Im[z]
  double yjumplo = 0.5 * (std::erf(z.real() - 
    1.5 * z.imag()) + 1.0);
  double yjumphi = 0.5 * (std::erf(z.real() + 
    1.5 * z.imag()) + 1.0);

  if (xi <= apprx_0_cdf) {
    if (yjumphi > 0.5 && yjumplo < 0.5) {
      jump *= (0.5 - yjumplo) / (yjumphi - yjumplo);
      yjumphi = 0.5;
    } else if (yjumplo > 0.5) {
      yjumplo = 0.0;
      yjumphi = 0.0;
      jump = 0.0;
    }
    if (jump > apprx_0_cdf) jump = apprx_0_cdf;
    const double d = yjumphi - yjumplo;
    const double sout = (apprx_0_cdf - jump) / (0.5 - d);
    const double sinv = jump > 0.0 ? d / jump : 0.0;
    if (xi >= sout * yjumplo + jump) {
      const auto r = sout * yjumplo + jump;
      const auto a = (r - dcdx * (yjumphi - 0.5) - apprx_0_cdf)
        / std::pow(yjumphi - 0.5, 2);
      return 0.5 * (-dcdx + std::sqrt(std::pow(dcdx, 2) - 
        4.0 * (apprx_0_cdf - xi) * a)) / a + 0.5;
    } else if (xi > sout * yjumplo) {
      return (xi - sout * yjumplo) * sinv + yjumplo;
    } else {
      if (sout > 0.0)
        return xi / sout;
      else return 0.5 * yjumplo;
    }
  } else { // xi > apprx_0_cdf
    if (yjumplo < 0.5 && yjumphi > 0.5) {
      jump *= (yjumphi - 0.5) / (yjumphi - yjumplo);
      yjumplo = 0.5;
    } else if (yjumphi < 0.5) {
      yjumplo = 1.0;
      yjumphi = 1.0;
      jump = 0.0;
    }

    // Clip innapropriately large jumps
    if (jump > 1.0 - apprx_0_cdf) jump = 1.0 - apprx_0_cdf;
    const auto d = yjumphi - yjumplo;
    const auto sout = (1.0 - jump - apprx_0_cdf) / (0.5 - d);
    const auto sinv = jump > 0.0 ? d / jump : 0.0;
    const auto thresh1 = sout * (yjumplo - 0.5) + jump + apprx_0_cdf;
    const auto thresh2 = sout * (yjumplo - 0.5) + apprx_0_cdf;
    
    if (xi >= thresh1)
      return (xi - thresh1) / sout + yjumphi;
    else if (xi > sout * (yjumplo - 0.5) + apprx_0_cdf)
      return (xi - thresh2) * sinv + yjumplo;
    else {
      const auto a = (thresh2 - dcdx * (yjumplo - 0.5) - 
        apprx_0_cdf)/std::pow(yjumplo - 0.5, 2);
      return 0.5 * (-dcdx + std::sqrt(std::pow(dcdx, 2) - 
      4.0 * (apprx_0_cdf - xi) * a)) / a + 0.5;
    }
  }

  UNREACHABLE();
}
\end{lstlisting}

\end{document}